\documentclass[11pt,fleqn]{article}
\usepackage{amsmath,amssymb,amsthm,cite,enumerate,graphics,epsfig,nth}
\usepackage{enumitem}

{ \theoremstyle{definition}

\newtheorem{remark}{Remark}
}

\newcommand{\todo}[1][\null]{\ensuremath{\clubsuit}}

\newcommand{\R}{{\mathbb R}}

\newcommand{\p}{\partial}

\newcommand{\Aut}{\mathop{\rm Aut}\nolimits}

\newcommand{\lsemioplus}{\mathbin{\mbox{$\lefteqn{\hspace{.77ex}\rule{.4pt}{1.2ex}}{\in}$}}}

\allowdisplaybreaks[1]

\begin{document}

%\title[Realizations of Galilei algebras]{Realizations of Galilei algebras}
\begin{center}
\LARGE \bf
Realizations of Galilei algebras
\end{center}

\begin{center}
Maryna Nesterenko$^\dag$, Severin Po\v sta$^{\ddag}$ and Olena Vaneeva$^{\dag}$
\end{center}

\noindent $^\dag$~Institute of Mathematics of NAS of Ukraine, 3
Tereshchenkivska Str., Kyiv-4, 01601 Ukraine\\
$\phantom{^\dag}$~E-mail: maryna@imath.kiev.ua, vaneeva@imath.kiev.ua

\noindent $^\ddag$~Department of Mathematics, Faculty of Nuclear Sciences and Physical Engineering,\\$\phantom{^\dag}$~Czech Technical University in Prague, Trojanova 13, CZ-120 00 Prague, Czech Republic\\
$\phantom{^\dag}$~E-mail: severin.posta@fjfi.cvut.cz

%\author{Maryna Nesterenko}
%\email{maryna@imath.kiev.ua}
%\affiliation{Institute of Mathematics of NAS of Ukraine, 3 Tereshchenkivska Str., 01601 Kyiv-4, Ukraine}
%\author{Severin Po\v sta}
%\email{severin.posta@fjfi.cvut.cz}
%\affiliation{Department of Mathematics, Faculty of Nuclear Sciences and Physical Engineering, Czech Technical University in Prague, Trojanova 13, CZ-120 00 Prague, Czech Republic}
%\author{Olena Vaneeva}
%\email{vaneeva@imath.kiev.ua}
%\affiliation{Institute of Mathematics of NAS of Ukraine, 3 Tereshchenkivska Str., 01601 Kyiv-4, Ukraine}

\date{\today}

\begin{abstract}
All inequivalent realizations of the Galilei algebras of dimensions not greater than five are constructed
using the algebraic approach proposed by I.~Shirokov.
The varieties of the deformed Galilei algebras are discussed and families of one-parametric deformations are presented in explicit form.
It is also shown that a number of well-known and physically interesting equations and systems are invariant with respect
to the considered Galilei algebras or their deformations.
\end{abstract}

%PACS numbers: 02.20Sv, 02.30.Jr, 02.30.Hq

%Key words: Galilei algebra, realization, left-invariant vector field, deformation of Lie algebra, Lie symmetry

\section{Introduction}
%The relativity principle is one of the corner stones of modern physics as the relevant theory of relativity plays the basic role in structuring a given physical theory. The relativity principle  was firstly introduced by Galileo Galilei.

A problem of description of all possible realizations of a Lie algebra by  first-order differential operators arises in different areas of mathematical physics and has a~number of practical applications (see~\cite{Popovych2003} for the references). In particular, realizations serve as prerequisite for construction of differential invariants and, consequently, for the invariant differential equations.
In this work we completely solve this problem for the five  Galilei algebras of lowest dimensions, namely for the reduced classical Galilei algebra ${\rm A\bar{G}}_1(1)$,
for the classical Galilei algebra ${\rm A{G}}_1(1)$,
for the reduced extended Galilei algebra ${\rm A\bar{G}}_2(1)$,
for the extended Galilei algebra ${\rm A{G}}_2(1)$
and
for the reduced special Galilei algebra ${\rm A\bar{G}}_3(1)$, see Section~\ref{sec_def_Galilei} for definitions.
The choice of Galilei algebra as the object of our investigation is motivated by the fact that
the Galilean invariance  principle underlies the classical mechanics, whose basic equations are invariant with respect to Galilei group of transformations of space and time variables. At the same time  the Galilei group and its extensions appear as symmetry groups of model equations not only in classical mechanics but in various fields of physics, e.g., Schr\"odinger and Levi-Leblond equations are invariant under the Galilei group. Moreover, Galilei symmetry of the Levi-Leblond equation allows one to predict the correct value of the gyromagnetic ratio. There also exist Galilei-invariant wave equations, which describe correctly the spin-orbit coupling~\cite{Nikitin}.
Mentioned applications require detailed knowledge of the structure of the Galilei group and the corresponding Galilei algebra~\cite{MR0466427}.

The direct method of construction of realizations can be successfully applied only for the particular Lie algebras, see, e.g., \cite{LahnoSpichakStognii2004}, in~\cite{Popovych2003} the constructive modification of direct approach was proposed, what made it possible to obtain all faithful realizations of the low-dimensional Lie algebras, but the proposed method leads to the cumbersome computations in the case of physically interesting algebras.
That is why to construct the realizations of Galilei algebras by first-order differential operators we use the connection between the realization and left-invariant vector field, and we make use of the algebraic approach to the construction of  vector fields proposed in~\cite{shirokov2013}, this method was already applied to the Poincar\'e algebras $p(1,1)$ and $p(1,2)$~\cite{Nesterenko2015}.
According to the Shirokov's method the necessary pre-condition to the construction of all realizations of a given Lie algebra is the description of its subalgebras up to inner and discrete automorphisms. For the Galilei algebras this task was partially solved in~\cite{Fushchych&Barannyk&Barannyk}, where subalgebras are classified up to the inner automorphisms. This classification can be enhanced once the complete groups of automorphisms are determined and the representatives equivalent with respect to the discrete automorphisms are eliminated.

Faithful realizations of several types of Galilei algebras were constructed previously, namely, the faithful realizations of ${\rm A\bar{G}}_1(1)$, ${\rm A{G}}_1(1)$ and  ${\rm A\bar{G}}_2(1)$ were derived for the isomorphic algebras of dimensions three and four in~\cite{Popovych2003}.
In this work we compare our results obtained by the Shirokov's method with those presented in~\cite{Popovych2003} and complete the lists by the unfaithful realizations and all realizations of the five-dimensional Galilei algebras.
Note that covariant realizations of the five-dimensional Galilei algebra ${\rm AG}_2(1)$ were constructed in~\cite{Rideau1993} using the direct method, and
later this classification was completed to the classification of all faithful realizations in~\cite{zhdanov1997a}.
Those results form  a subcase of the classification presented in Section~\ref{sec_AG21}, where  the unfaithful realizations are also constructed.
The work~\cite{Rideau1993} (resp.~\cite{zhdanov1997a}) also contains the covariant (resp. faithful) realizations of four-dimensional algebra ${\rm AG}_1(1)$ and six-dimensional algebra ${\rm AG}_3(1)$. The six-dimensional case is not considered here and four-dimensional case is a subject of Section~\ref{ssec_def_AG_11} and already was considered in~\cite{Popovych2003}.
Some covariant realizations were also obtained in~\cite{Lahno1998}, but  the realizations were additionally restricted by the fixed type of equation. Covariant realizations of the classical Galilei algebra of four-dimensional space-time are presented in~\cite[Appendix~3]{zhdanov1997}, see also references therein.

The structure of this paper is as follows.
We start with the basic definitions and description of the algorithm that we use for complete classification of realizations of low-dimensional Galilei algebras.
Then, in Section~\ref{sec_def_Galilei} we fix the notations that we use for the identification of different types of Galilei algebras.
The main result of the paper is contained in Sections~\ref{sec_34d} and~\ref{sec_5d},
where each of the Galilei algebras of dimensions three, four and five are considered together with the lists of inequivalent subalgebras and their complementary spaces, the respective realizations are constructed and discussed.

Section~\ref{sec_deformations} is devoted to the deformations of Galilei algebras and to the generic realizations of the deformed algebras. To construct explicit one-parametric families of the deformed algebras we used the known contractions of low-dimensional Lie algebras~\cite{Nesterenko1},
the known physical contractions between classical Lie algebras and classification lists for Lie algebras of fixed dimensions.

It turned out that a number of well-known differential equations with straight physical connections, such as Korteweg--de Vries and Kawahara equations, reaction--diffusion equations, the Ermakov system, central force and Kepler problems, are invariant with respect to the low-dimensional Galilei algebras or their deformations.
In Section~\ref{sec:galinvdifeq} we present explicitly Galilei-invariant equations with the respective realizations and subalgebras.

\section{Construction of realizations}\label{sec_theory}

Let $\mathfrak g$ be an $n$-dimensional Lie algebra spanned by a basis
$\{e_1, e_2,\dots, e_n\}$ with the structure constants $C_{ij}^k\in\mathbb{R}$, where $i,j,k=1,2,\dots,n$.
We denote an open domain of $\mathbb{R}^m$ as $M$ and the Lie algebra of vector fields on it as ${\rm Vect}(M)$.
%,
%%the vector fields are taken in a form of linear first-order differential operators with analytical coefficients
%and the Lie product of vector fields is given by their commutator.
The automorphism group of $\mathfrak g$ is denoted by ${\rm Aut}(\mathfrak g)$.
%and the notion of realization of a Lie algebra is defined as follows.

%\begin{definition}
A \emph{realization} of the Lie algebra $\mathfrak g$ by vector fields on $M$ is a homomorphism $R\colon \mathfrak g \rightarrow {\rm Vect}(M).$
The~realization is {\it faithful} if $\mathop{\rm ker}\nolimits R=\{0\}$ and {\it unfaithful} otherwise.
%\end{definition}

Let $x=(x_1,\dots,x_m)$ be the coordinates on $M$ and $X_i\in {\rm Vect}(M)$ be the images of the basis elements $e_i$ of
$\mathfrak g$ under the realization $R$, i.e.,
$X_i=R(e_i)=\xi^i_a(x)\p_a$, where $a=1,\ldots,m$, $\p_a=\tfrac{\p}{\p_{x_a}}$,
and $a$ is the summation index.
Denote the local transformation group that corresponds to the vector fields $X_i$  as $G$,
that is, $X_i$ are  the infinitesimal generators of the action of $G$ on $M$.

%\begin{definition}
If the group $G$ acts on $M$ regularly, i.e., the action of~$G$ is  both transitive and free,
then the corresponding realization is called \emph{generic}.
For any generic realization the dimension of $M$ coincides with the dimension of~$\mathfrak g$, that is $m=n$.
%\end{definition}

Below we present the practical algorithm for construction of left-invariant vector fields that is based on results of~\cite{shirokov1997, shirokov2013},
then the basis elements $e_i$ of the Lie algebra~$\mathfrak g$ are realized by linearly independent left-invariant vector fields~$\xi_i$.
Since the components of the left-invariant vector fields $\xi_i(x)=\xi^i_k(x)\partial_{x_k}$
always can be recovered from the left-invariant differential one-forms $\omega^j(x)=\omega^j_i(x)dx_j$ using the duality
$\omega^j_i(x)\xi^i_k(x)=\delta^j_k$, so our problem is reduced to the determination of the components of $\omega^j$.

Let the adjoint action of~$G$ is regular, we fix second canonical coordinates $g_1(x_1), \dots$, $g_n(x_n)$ on~$G$ associated with the basis $e_1,\dots,e_n$,
then the components $\omega^j_i(x)$ of the left-invariant differential one-forms are computed as follows:
\begin{equation*}
\omega^j_i(x)=\left(A_{1}(x_1)\cdots A_{i-1}(x_{i-1})\right)^j_i,\quad
%i=2,\dots, n,\quad
%j=1,\dots, n,\quad
\omega^j_1=\delta^j_1.
\end{equation*}
Here each of the matrices $A_{i}$ is the  solution of the initial value problem
\[\dot{A}_{i}(t)=-{\rm ad}_{e_i} A_{i}(t),\quad
A_{i}(0)=I,
\]
where $\dot{A}_{i}(t)$ is the derivative matrix, therefore,
\[\omega^j_i(x)=\left({\rm exp}{(-x_1{\rm ad}_{e_1})}\cdots {\rm exp}{(-x_{i-1}{\rm ad}_{e_{i-1}})}\right)^j_i, \quad ({\rm ad}_{e_k})_i^j=C_{ki}^j.\]

Therefore we present the construction of the left-invariant vector fields, what gives the generic realization of the algebra $\mathfrak g$.
All the rest of transitive realizations are constructed by means of projection of the generic realization using the following rule.

Let $H$ be the isotropy subgroup of a fixed point $x_0\in M$ with the corresponding Lie algebra~$\mathfrak h$.
Then $G$ acts transitively on the space of right cosets $H{\setminus} G$ and we can identify $M$ with $H{\setminus} G$ and introduce the induced coordinates $y=(x, h)$, where $h\in H$ and $x\in M$ (actually $H{\setminus} G$).
Let second canonical coordinates are associated with the basis $e_1,\dots, e_n$ in such a way that $\mathfrak h=\langle e_{m+1},\dots,e_n\rangle$ and the space complementary to $\mathfrak h$ is spanned by the elements $\{e_1,\dots,e_m\}$. Then the left-invariant vector fields on $G$ have the form
\begin{align*}
&\xi_i=\xi_{i}^1(x)\partial_1+\dots+\xi_{i}^m(x)\partial_m+
\xi_{i}^{m+1}(x, h)\partial_{m+1}+\dots+\xi_{i}^n(x,h)\partial_n,
\quad i=1,\dots, n,
\end{align*}
and in the chosen coordinates the generators of the group action are obtained by the projection
\begin{equation*}
\textrm{pr}_{\mathfrak h}(\xi_i)=X_i=\xi_{i}^1(x_1,\dots, x_m)\partial_1+\dots+\xi_{i}^m(x_1,\dots, x_m)\partial_m.
\end{equation*}

Therefore, for each subalgebra $\mathfrak h$ of $\mathfrak g$ we can construct the realization $R(e_i)=\textrm{pr}_{\mathfrak h}(\xi_i)$
and all possible realizations are exhausted by those that correspond to subalgebras inequivalent with respect to inner and discrete automorphisms.

The number of variables necessary to realize the Lie algebra is
$m=\dim(\mathfrak{g})-\dim(\mathfrak{h})$ and it coincides with the codimension of the subalgebra~$\mathfrak{h}$
and the projected realization $R_{\mathfrak{h}}(\mathfrak{g})$ is called a \emph{realization respective to the subalgebra}~$\mathfrak h$.
In~particular, this means that the generic realization corresponds to the zero subalgebra $\mathfrak h_{0}=\langle 0\rangle$
and $R_{\mathfrak{h}_0}(\mathfrak{g})$ is always realized in $n=\dim(\mathfrak{g})$ variables.
Further in this paper we do not include $\mathfrak{h}_0$ to the list of inequivalent subalgebras, but a generic realization $R_{\mathfrak{h}_0}$
is constructed for each Lie algebra.

Suppose that instead of the fixed subalgebra $\mathfrak{h}$
%spanned by the basis elements $e_{m+1},\dots,e_n$
we consider a set of subalgebras $\mathfrak{h}^\alpha$ smoothly parametrized by the parameter $\alpha\in \R$, then for each fixed value $\alpha_0$ we can apply the above algorithm and obtain the realization $R_{\mathfrak{h}^{\alpha_0}}$.
The obtained set of realizations can be interpreted in two ways~\cite{Shirokov_letter}
\begin{itemize}\itemsep=0pt
\item
we constructed the $\R$-parametrized series $R_{\mathfrak{h}^{\alpha}}$ of realizations;

\item
the parameter $\alpha$ is a function of independent invariants  of the transformation group $G$, but these invariants are  chosen as a part of local coordinates, namely $x_{m+1},x_{m+2}, \dots, x_n$, i.e., $\alpha=\alpha(x_{m+1},x_{m+2}, \dots, x_n)$ and without loss of generality we can put $\alpha=x_{m+1}$; therefore, we have the realization $R_{\mathfrak{h}^{x_{m+1}}}$ in $(m+1)$ variables.
\end{itemize}
Usually the range of variation of the parameter is not equal to $\R$, what is caused by the equivalence of subalgebras for some different values of parameter, then the same range of variation should be preserved for $\alpha$ in the parametrized realization and for  $x_{m+1}$ in the realization with the additional variable.
If the subalgebra is parametrized by more than one parameter, then the above procedure should be carried out for each parameter independently.

%\begin{remark}
The considered algorithm produces polynomial coefficients of the partial
differential operators only in the case of nilpotency of the adjoint representations of basis elements.

The structure of realizations constructed by means of the algebraic method reminds a tree diagram, namely:
a realization respective to a subalgebra $\mathfrak{h}_1$  can be constructed by means of projection from
a realization respective to a subalgebra $\mathfrak{h}_2$ if $\mathfrak{h}_2\subset \mathfrak{h}_1$.
%\end{remark}

One more specific property of the algorithm is that the first basis element of the space complementary to the subalgebra
is represented by  the shift (translation) operator, and, in the case if the second basis element of the complementary space commutes with the first one, then the second operator is also a shift one, etc.
Therefore, if it is known from the application which of the basis elements are to be the shift operators,
then these elements should be posed (re-enumerated) on the first places before the algorithm application.

In this paper we purposely do not fix independent (say $t$, $x$) and dependent (say $u$, $w$)
variables in the presented realizations. One can chose them in any order convenient for the certain application.
For example, the same realization can be considered as
\begin{alignat*}{3}
&\{\partial_t,\; \partial_x,\; x\partial_t+t\partial_x+\partial_u\}\quad &&\text{for} \ u=u(t,x) \quad \text{or}&\\
&\{\partial_u,\, \partial_w,\, w\partial_u+u\partial_w+\partial_t\}\quad &&\text{for} \ u=u(t) \ \text{and} \ w=w(t).&
\end{alignat*}

\section{Galilei algebras: definitions and conventions}\label{sec_def_Galilei}

In this section we give definitions of main types of Galilei Lie algebras and their basis operators.
For all Lie algebras considered in this paper we present the nonzero commutation relations only.
First we consider the most general case.
\emph{The full Galilei algebra} ${\rm AG}_4(n)$ of $(n+1)$-dimensional Newton space-time is generated by the basis elements
\begin{enumerate}[label=\arabic*)]\itemsep=0pt
%\begin{enumerate}
\item
$P_i$ (operators of spatial translations, $i=1,2,\dots, n$);
\item
$T$ (operator of temporal translation);
\item
$J_{ij}$ (operators of rotations, $i<j$, $i,j=1,2,\dots, n$);
\item
$G_i$ (operators of pure Galilei transformations, $i=1,2,\dots, n$);
\item
$D$ (dilatation operator);
\item
$S$ (projection operator);
\item
$Z$ (operator of scale transformation of spatial variables);
\item
$M$ (mass operator).
\end{enumerate}

These basis elements are connected by the commutation relations:
\begin{gather}
[J_{ij}, J_{kl}] =\delta_{il} J_{jk} +\delta_{jk} J_{il} - \delta _{ik} J_{jl} -\delta_{jl} J_{ik},
\quad i,j,k,l=1,2,\dots, n;
\label{comm_rel_JJJ}
\\
[P_i, J_{jk}] =\delta_{ij} P_k -\delta_{ik} P_j;\label{comm_rel_PJP}
\\
[T, G_i] = -P_i;\label{comm_rel_TGP}
\\
[G_i, J_{jk}] =\delta_{ij} G_k -\delta_{ik} G_j;\label{comm_rel_GJG}
\\
[D, G_i] =G_i;\label{comm_rel_DGG}
\\
[D, P_i] =-P_i;\label{comm_rel_DPP}
\\
[D, T] =-2T;\label{comm_rel_DTT}
\\
[S, P_i] =G_i;\label{comm_rel_SPG}
\\
[D, S] =2S;\label{comm_rel_DSS}
\\
[T, S] =D;\label{comm_rel_TSD}
\\
[Z, G_i] =-G_i;\label{comm_rel_ZGG}
\\
[Z, P_i] =-P_i;\label{comm_rel_ZPP}
\\
[G_i, P_j] =\delta_{ij} M;\label{comm_rel_GPM}
\\
[Z, M] =-2M.\label{comm_rel_ZMM}
\end{gather}

The full Galilei algebra contains several subalgebras and reductions which are important in physics and all of them are usually called Galilei algebra,
we define them precisely in Table~\ref{tab:galalgbrs}.

\begin{table}[b!]\small\centering
\begin{tabular}{lllll}
\hline
&&&&\\[-8pt]
\hfil Galilei~algebra & Notation & Basis elements &\parbox{2 cm}{Commutation\\ relations}& Dimension  \\[8pt]
\hline
&&&&\\[-8pt]
\parbox{2.7 cm}{\emph{Classical}}&${\rm AG}_1(n)$&
$\langle P_i,\, T,\, J_{ij},\, G_i,\, M\rangle$&(\ref{comm_rel_JJJ})--(\ref{comm_rel_GJG}), (\ref{comm_rel_GPM})&$\tfrac{1}{2}n(n-1)+2n+2$
\\[8pt]
\parbox{2.7 cm}{\emph{Extended}}&${\rm AG}_2(n)$&
$\langle P_i,\, T,\, J_{ij},\, G_i,\, D,\, M\rangle$&(\ref{comm_rel_JJJ})--(\ref{comm_rel_DTT}), (\ref{comm_rel_GPM})&$\tfrac{1}{2}n(n-1)+2n+3$
\\[8pt]
\parbox{2.7 cm}{\emph{Special}}&${\rm AG}_3(n)$&
$\langle P_i,\, T,\, J_{ij},\, G_i,\, D,\, S,\, M\rangle$&(\ref{comm_rel_JJJ})--(\ref{comm_rel_TSD}),  (\ref{comm_rel_GPM})&$\tfrac{1}{2}n(n-1)+2n+4$
\\[8pt]
\parbox{2.5 cm}{\emph{Full}}&${\rm AG}_4(n)$&
$\langle P_i,\, T,\, J_{ij},\, G_i,\, D,\, S,\, Z,\, M\rangle$&(\ref{comm_rel_JJJ})--(\ref{comm_rel_ZMM})&$\tfrac{1}{2}n(n-1)+2n+5$
\\[8pt]
\parbox{2.7 cm}{\emph{Reduced~classical}}&${\rm A\bar{G}}_1(n)$&
$\langle P_i,\, T,\, J_{ij},\, G_i\rangle$&(\ref{comm_rel_JJJ})--(\ref{comm_rel_GJG})&$\tfrac{1}{2}n(n-1)+2n+1$
\\[8pt]
\parbox{2.7 cm}{\emph{Reduced~extended}}&${\rm A\bar{G}}_2(n)$&
$\langle P_i,\, T,\, J_{ij},\, G_i,\, D\rangle$&(\ref{comm_rel_JJJ})--(\ref{comm_rel_DTT})&$\tfrac{1}{2}n(n-1)+2n+2$
\\[8pt]
\parbox{2.7 cm}{\emph{Reduced~special}}&${\rm A\bar{G}}_3(n)$&
$\langle P_i,\, T,\, J_{ij},\, G_i,\, D,\, S\rangle$&(\ref{comm_rel_JJJ})--(\ref{comm_rel_TSD})&$\tfrac{1}{2}n(n-1)+2n+3$
\\[8pt]
\parbox{2.7 cm}{\emph{Reduced~full}}&${\rm A\bar{G}}_4(n)$&
$\langle P_i,\, T,\, J_{ij},\, G_i,\, D,\, S,\, Z\rangle$&(\ref{comm_rel_JJJ})--(\ref{comm_rel_ZPP})&$\tfrac{1}{2}n(n-1)+2n+4$
\\[8pt]
\hline
\end{tabular}
\caption{Types of Galilei algebras.}\label{tab:galalgbrs}
 \end{table}

\section{Three- and four-dimensional cases}\label{sec_34d}
If we look for the lowest possible Galilei algebras, then from the previous section it can be concluded that there exist only one three-dimensional and two four-dimensional Galilei algebras:
\begin{itemize}\itemsep=0pt
\item
the three-dimensional Lie algebra of the reduced  classical Galilei group of one-dimensional space ${\rm A\bar{G}}_1(1)$;
\item
the four-dimensional Lie algebra of the classical Galilei group $A G_1(1)$ with a mass operator~$M$;
\item
the four-dimensional Lie algebra of the reduced extended Galilei group  of one-dimensional space ${\rm A\bar{G}}_2(1)$.
\end{itemize}

\subsection{The reduced classical Galilei algebra $\boldsymbol{{\rm A\bar{G}}_1(1)}$}
The reduced classical Galilei algebra ${\rm A\bar G}_1(1)$ of one-dimensional space is generated by the operators $P$, $G$ and $T$
with the nonzero commutation relation $[G,T]=P$.

As a Lie algebra ${\rm A\bar{G}}_1(1)$ coincides with the real three dimensional Lie algebra $A_{3.1}$
(according to the Mubarakzyanov's~\cite{mubarakzyanov1963.1} classification) with the unique nonzero commutator $[e_2,e_3]=e_1$.

It was proven in~\cite{Popovych2003} that all inequivalent faithful realizations of $A_{3.1}$ are exhausted by the following list
(we denote the partial derivatives as  $\p_i=\frac{\p}{\p_{x_i}}$)
\begin{gather*}
e_1=\p_1,\quad e_2=\p_2,\quad e_3=x_2\p_1+\p_3;\\
e_1=\p_1,\quad e_2=\p_2,\quad e_3=x_2\p_1+x_3\p_2;\\
e_1=\p_1,\quad e_2=\p_2,\quad e_3=x_2\p_1.
\end{gather*}

The full automorphism group $\Aut({\rm A\bar{G}}_1(1))$ is generated by the nonsingular matrices
\begin{equation*}
\begin{pmatrix}
\alpha_{22}\alpha_{33}-\alpha_{23}\alpha_{32} & 0 & 0  \\
\alpha_{21} & \alpha_{22} & \alpha_{23}  \\
\alpha_{31}& \alpha_{32} & \alpha_{33}
\end{pmatrix},\quad \mbox{where}\ \alpha_{ij} \in \R,\ \mbox{and}\ \alpha_{22}\alpha_{33}\neq\alpha_{23}\alpha_{32}.
\end{equation*}

Everywhere in this paper we suppose that automorphism matrix acts from the left and the order of basis elements coincides with the order indicated in a Lie algebra definition.

All subalgebras of ${\rm A\bar{G}}_1(1)$ which are inequivalent with respect to inner automorphisms are
\begin{gather*}
\langle 0\rangle,\
\langle P\rangle,\
\langle G\rangle,\
\langle T\rangle,\
\langle T+\alpha G\rangle,\
\langle P, \,G\rangle,\
\langle P, \,T\rangle,\
\langle P,\,T+\alpha G\rangle,\
\langle P, \,G,\, T\rangle\ (\alpha>0).
\end{gather*}

The discrete external automorphism $\tilde G=T$, $\tilde T=G$, $\tilde P=-P$,
whose matrix of $\Aut({\rm A\bar{G}}_1(1))$ is of the form
$\left(\begin{smallmatrix}
-1&0&0\\
0&0&1\\
0&1&0
\end{smallmatrix}\right)$,
implies equivalence  of subalgebras in the pairs $(\langle T \rangle,\, \langle G\rangle)$ and $(\langle P,\, T\rangle,\,\langle P, \,G\rangle)$.

The rest of subalgebras are inequivalent with respect to  automorphisms and we label each inequivalent subalgebra $\mathfrak h_{k.p}$ by its dimension~`$k$' and (if necessary) it's number~`$p$' in the list. In some cases we also combine subalgebras by means of extension of the parameter value,
e.g., $\langle T \rangle$ and $\langle T+\alpha G \rangle$, $\alpha>0$ are transformed into $\langle T+\alpha G \rangle$, $\alpha\ge 0$. Thus,   proper inequivalent subalgebras  of ${\rm A\bar{G}}_1(1)$ are
\begin{gather*}
\mathfrak h_{1.1}= \langle P \rangle;
\quad
\mathfrak h^\alpha_{1.2}= \langle T+\alpha G \rangle\ (\alpha\ge 0);
\quad
\mathfrak h^\alpha_{2}= \langle P, \, T+\alpha G \rangle\ (\alpha\ge 0).
\end{gather*}

Note that ideals of ${\rm A\bar{G}}_1(1)$ are $\langle P \rangle$ and $\langle P,\, pG+qT \rangle$, $p,q \in \R$,
therefore the subalgebras ${\mathfrak h}_{1.1}$, ${\mathfrak h}_{2}$~and~${\mathfrak h}_{3}$
contain ideals and, hence, lead to unfaithful realizations, which were not obtained in~\cite{Popovych2003}.

In Table~\ref{tab:realredAg11} we list realizations $R_{{\mathfrak h}_{k.p}}$ of ${\rm A\bar{G}}_1(1)$ corresponding to subalgebras $\mathfrak h_{k.p}$,
before each realization we adduce the vector space complementary to $\mathfrak h_{k.p}$ that was used to obtain the corresponding  realization, the same notations will be used in other tables.
\begin{table*}\small
\renewcommand{\arraystretch}{1.1}\centering
%\begin{ruledtabular}
\begin{tabular}{rl}
\hline
&\\[-8pt]
\parbox{2.7 cm}{Complementary\\ basis} & Realization\\[8pt]
\hline
$\{P,T,G\}$&$R(\mathfrak h_{0})\colon\   P=\partial_1,\  G=-x_2\partial_1+\partial_3,\  T=\partial_2$
\\
$\{G,T\}$&
$R_{\mathfrak h_{1.1}}\colon\   P=0,\ G=\partial_1,\  T=\partial_2$
\\
$\{P, G\}$&
$R_{\mathfrak h^\alpha_{1.2}}\colon \   P=\partial_1,\  G=\partial_2,\   T=x_2\partial_1-\alpha\partial_2$
\\
&$R_{\mathfrak h^{x_3}_{1.2}}\colon \   P=\partial_1,\   G=\partial_2,\  T=x_2\partial_1-{x_3}\partial_2$
\\
$\{G\}$& $R_{\mathfrak h^{\alpha}_{2}}\colon \    P=0,\  G=\partial_1,\    T=-\alpha\partial_1$
\\
& $R_{\mathfrak h^{x_2}_{2}}\colon \    P=0,\  G=\partial_1,\    T=-x_2\partial_1$
\\[4pt]
\hline
\end{tabular}
\caption{Realizations of the reduced classical Galilei algebra ${{\rm A\bar{G}}_1(1)}$.}
\label{tab:realredAg11}
%\end{ruledtabular}
\end{table*}

As far as subalgebras $h^\alpha_{1.2}$ and $h^\alpha_{2}$ are parametrized then we have two more realizations
\begin{alignat*}{5}
&R_{\mathfrak h^{x_3}_{1.2}}\colon\  && P=\partial_1,\quad &&  G=\partial_2,\quad && T=x_2\partial_1-{x_3}\partial_2;& \\
& R_{\mathfrak h^{{x_2}}_{2}}\colon\  && P=0,\quad && G=\partial_1,\quad && T=-{x_2}\partial_1.&
\end{alignat*}

\begin{remark}
The realizations $R_{{\mathfrak h}_{0}}$, $R_{{\mathfrak h}^0_{1.2}}$ and $R_{{\mathfrak h}^{x_3}_{1.2}}$ of the algebra ${\rm A\bar{G}}_1(1)$ coincide with those obtained in \cite{Popovych2003},
the rest of realizations are new.

The realization  $R_{{\mathfrak h}^\alpha_{1.2}}$ for arbitrary $\alpha>0$ can be transformed to  $R_{{\mathfrak h}^0_{1.2}}$ by means of the automorphism transformation
$\left(\begin{smallmatrix}
1&0&0\\
0&1&0\\
0&\alpha&1
\end{smallmatrix}\right).$
\end{remark}

\subsection{The classical Galilei algebra $\boldsymbol{{\rm AG}_1(1)}$}
The classical Galilei
algebra ${\rm AG}_1(1)$ is generated by the four operators $P$, $G$, $T$ and~$M$,
that satisfy nonzero commutation relations $[G,T]=P$ and $[G,P]=M$.
${\rm AG}_1(1)$ is isomorphic to the Lie algebra $A_{4.1}$ with the commutators $[e_2,e_4]=e_1$ and $[e_3,e_4]=e_2$,
the isomorphism is established by the following relations{\samepage
\begin{equation}\label{baseAG11}
P=e_2,\quad T=e_3,\quad G=-e_4,\quad M=e_1.
\end{equation}
Inequivalent faithful realizations of $A_{4.1}$ are given in Table~\ref{tab:faithfreala41}.}

\begin{table*}\small
\renewcommand{\arraystretch}{1.1}
\centering
\begin{tabular}{lrlllll}
\hline
& 1) & $e_1=\p_1$,& $e_2=\p_2$,& $e_3=\p_3$, & $e_4=x_2\p_1+x_3\p_2+\p_4$&\\
& 2) & $e_1=\p_1$,& $e_2=\p_2$, & $e_3=\p_3$, &  $e_4=x_2\p_1+x_3\p_2+x_4\p_3$ &\\
& 3) & $e_1=\p_1$, & $e_2=\p_2$, & $e_3=\p_3$,& $e_4=x_2\p_1+x_3\p_2$&\\
& 4) & $e_1=\p_1$ &  $e_2=\p_2$, & $e_3=x_3\p_1+x_4\p_2$,&  $e_4=x_2\p_1+x_4\p_3-\p_4$&\\
& 5) & $e_1=\p_1$, &  $e_2=\p_2$, &  $e_3=-\tfrac 12 x_3^2 \p_1+x_3\p_2$, & $e_4=x_2\p_1-\p_3$&\\
& 6) & $e_1=\p_1$, &  $e_2=x_2\p_1$, & $e_3=\p_3$, &  $e_4=x_2x_3\p_1-\p_2$&\\
& 7) & $e_1=\p_1$, & $e_2=x_2\p_1$, & $e_3=x_3\p_1$, & $e_4=-\p_2-x_2\p_3$&\\
& 8) & $e_1=\p_1$, &  $e_2=x_2\p_1$, &  $e_3=\tfrac 12 x_2^2 \p_1$, & $e_4=-\p_2$&\\
\hline
\end{tabular}
\caption{Inequivalent faithful realizations of $A_{4.1}$.}
\label{tab:faithfreala41}
\end{table*}

The full group of automorphisms of ${\rm AG}_1(1)=\langle M,\, P,\, T,\, G\rangle$ is generated by the nonsingular matrices
\begin{equation*}
\begin{pmatrix}
\alpha_{33}\alpha_{44}^2& 0 & 0& 0 \\
\alpha_{32}\alpha_{44}& \alpha_{33}\alpha_{44} &0&0\\
\alpha_{31}&\alpha_{32}&\alpha_{33}&0\\
\alpha_{41}& \alpha_{42} &\alpha_{43} & \alpha_{44}
\end{pmatrix},\quad \mbox{where}\ \alpha_{ij} \in \R,\ \mbox{and}\ \alpha_{33}\alpha_{44}\neq0.
\end{equation*}

Using the results of~\cite{Fushchych&Barannyk&Barannyk} and automorphisms of ${\rm AG}_1(1)$ we obtain the following list of inequivalent subalgebras
\begin{alignat*}{4}
%&\mathfrak h_{0}= \langle 0 \rangle;
%\\
&\mathfrak h_{1.1}= \langle P \rangle, \quad &&\mathfrak h_{2.1}= \langle M,\,P \rangle,\quad && \mathfrak h_{3.1}= \langle M,\,P,\,G \rangle,&\\
&\mathfrak h^\alpha_{1.2}= \langle \alpha T+G \rangle\ (\alpha\ge 0), \quad &&\mathfrak h_{2.2}= \langle M,\,G \rangle,\quad &&
\mathfrak h^\alpha_{3.2}= \langle M, \,P,\,T+\alpha G \rangle\ (\alpha\ge 0).&\\
&\mathfrak h^\beta_{1.3}= \langle  T+\beta M \rangle \ (\beta\in\mathbb R),\quad && \mathfrak h^\alpha_{2.3}= \langle M, \,T+\alpha G \rangle\ (\alpha\ge 0),\quad &&&\\
&\mathfrak h_{1.4}= \langle M \rangle;\quad && \mathfrak h^\beta_{2.4}= \langle P, \,T+\beta M \rangle \ (\beta\in\mathbb R); \quad &&&
\end{alignat*}

Ideals of ${\rm AG}_1(1)$ are $\langle M \rangle$, $\langle M, \,P \rangle$, $\langle M, \,P, \,T \rangle$ and
$\langle M, \,P, \,T+\alpha G\rangle$, therefore the realizations corresponding to subalgebras
$\mathfrak h_{1.4}$, $\mathfrak h_{2.1}$, $\mathfrak h_{2.2}$, $\mathfrak h^\alpha_{2.3}$, $\mathfrak h_{3.1}$, $\mathfrak h^\alpha_{3.2}$
will be unfaithful.

The obtained realizations of ${\rm AG}_1(1)$ are listed in Table~\ref{tab:realag11b}.

\begin{table*}\small
\renewcommand{\arraystretch}{1.1}
\centering
\begin{tabular}{rl}
\hline
&\\[-8pt]
\parbox{2.7 cm}{Complementary\\ basis} & Realization\\[8pt]
\hline
&\\[-8pt]
$\{ M,\,P,\,T,\,G\}$ &$  R_{\mathfrak h_{0}}\colon\   M=\p_1,\    P=-\p_2,\   T=\p_3,\   G=x_2\p_1+x_3\p_2+\p_4$
\\
$\{ M,\,T,\,G\}$ &$  R_{\mathfrak h_{1.1}}\colon\   M=\p_1,\    P=x_3\p_1,\   T=\p_2,\ G=-x_2x_3\p_1+\p_3$
\\
$\{ M,\,P,\,T\}$ &$  R_{\mathfrak h^\alpha_{1.2}}\colon\   M=\p_1,\    P=\p_2,\   T=\p_3,\  G=-x_2\p_1-x_3\p_2-\alpha\p_3$
\\
&$R_{\mathfrak h^{x_4}_{1.2}}\colon\     M=\p_1,\    P=\p_2,\   T=\p_3,\  G=-x_2\p_1-x_3\p_2-x_4\p_3$
\\
$\{ M,\,P,\,G\}$ &$  R_{\mathfrak h^\beta_{1.3}}\colon\   M=\p_1,\    P=-\p_2,\   T=-\left(\tfrac12x_3^2+\beta\right)\p_1-x_3\p_2,\   G=x_2\p_1+\p_3$
\\
&$R_{\mathfrak h^{x_4}_{1.3}}\colon\  M=\p_1,\    P=-\p_2,\   T=-\left(\tfrac12x_3^2+x_4\right)\p_1-x_3\p_2,\  G=x_2\p_1+\p_3$
\\
$\{ P,\,T,\,G\}$ &$  R_{\mathfrak h_{1.4}}\colon\   M=0,\    P=-\p_1,\  T=\p_2,\   G=x_2\p_1+\p_3$
\\
$\{ T,\,G\}$&$  R_{\mathfrak h_{2.1}}\colon\   M=0,\    P=0,\   T=\p_1,\ G=\p_2$
\\
$\{ P,\,T\}$&$  R_{\mathfrak h_{2.2}}\colon\   M=0,\    P=-\p_1,\   T=\p_2,\ G=x_2\p_1$
\\
$\{ P,\,G\}$&$  R_{\mathfrak h^\alpha_{2.3}}\colon\   M=0,\    P=-\p_1,\   T=-x_2\p_1-\alpha\p_2,\ G=\p_2$
\\
&$R_{\mathfrak h^{x_3}_{2.3}}\colon\   M=0,\    P=-\p_1,\   T=-x_2\p_1-x_3\p_2,\   G=\p_2$
\\
$\{ M,\,G\}$&$  R_{\mathfrak h^\beta_{2.4}}\colon\   M=\p_1,\    P=x_2\p_1,\   T=\left(\tfrac12x_2^2-\beta\right)\p_1,\ G=\p_2$
\\
&$R_{\mathfrak h^{x_3}_{2.4}}\colon\   M=\p_1,\    P=x_2\p_1,\   T=\left(\tfrac12x_2^2-x_3\right)\p_1,\  G=\p_2$
\\
$\{ T\}$&$  R_{\mathfrak h_{3.1}}\colon\   M=0,\    P=0,\   T=\p_1,\ G=0$
\\
$\{ G\}$&$  R_{\mathfrak h^\alpha_{3.2}}\colon\  M=0,\    P=0,\   T=-\alpha\p_1,\ G=\p_1$
\\
&$R_{\mathfrak h^{x_2}_{3.2}}\colon\   M=0,\    P=0,\   T=-x_2\p_1,\   G=\p_1$
\\[4pt]
\hline
\end{tabular}
\caption{Realizations of the classical Galilei algebra ${{\rm AG}_1(1)}$.}
\label{tab:realag11b}
\end{table*}

The parametrized series of  realizations $R_{{\mathfrak h}^{\alpha}_{1.2}}$ can be transformed to the unique form
\begin{gather*}
R\big(\mathfrak h^0_{1.2}\big)\colon\  M=\p_1,\quad   P=\p_2,\quad  T=\p_3,\quad  G=-x_2\p_1-x_3\p_2
\end{gather*}
via the automorphism transformation
$\left(\begin{smallmatrix}
1 & 0& 0& 0\\
0 & 1 & 0& 0\\
0&0 &1&0\\
0 & 0 & \alpha & 1\\
\end{smallmatrix}\right)$.

{\samepage Another parametrized series of  realizations $R(\mathfrak h^{\beta\ne0}_{1.3})$ can be transformed to the unique form
\begin{gather*}
R\big(\mathfrak h^0_{1.3}\big)\colon\ M=\p_1,\quad   P=\p_2,\quad T=-\tfrac12x_3^2\p_1+x_3\p_2,\quad G=-x_2\p_1-\p_3
\end{gather*}
by means of the change of variables $-x_2\mapsto x_2$ and the automorphism transformation
$\left(\begin{smallmatrix}
1 & 0& 0& 0\\
0 & 1 & 0& 0\\
\beta&0 &1&0\\
0 & 0 & 0 & 1\\
\end{smallmatrix}\right)$.}

The same automorphism transforms the parametrized series of  realizations $R(\mathfrak h^{\beta}_{2.4})$ to the unique form
\begin{gather*}R\big(\mathfrak h^0_{2.4}\big)\colon\ M=\p_1,\quad   P=x_2\p_1,\quad T=\tfrac12x_2^2\p_1,\quad   G=\p_2.\end{gather*}

Using the nonsingular change of variables $\tfrac 12 x_2-x_3\mapsto x_3$ we transform the realization $R_{{\mathfrak h}^{x_3}_{2.4}}$ to \begin{gather*}M=\p_1,\quad P=x_2\p_1,\quad T=x_3\p_1,\quad G=\p_2+x_2\p_3.\end{gather*}

Application of the transformations $x_2\mapsto-x_2$, $-(\tfrac 12 x_3^2+x_4)\mapsto x_3$ and $x_3\mapsto x_4$
to the realization $R_{{\mathfrak h}^{x_4}_{1.3}}$
 results in the following form of operators
\begin{gather*}
M=\p_1,\quad   P=x_2\p_1,\quad T=x_3\p_1+x_4\p_2,\quad  G=-x_2\p_1+x_4\p_3+\p_4.
\end{gather*}

Now it is obvious that the respective realizations of $A_{4.1}$  and ${\rm AG}_1(1)$ are equivalent
$1)\sim R_{{\mathfrak h}_0}$,
$2)\sim R_{{\mathfrak h}^{x_4}_{1.2}}$,
$3)\sim R_{{\mathfrak h}^0_{1.2}}$,
$4)\sim R_{{\mathfrak h}^{x_4}_{1.3}}$,
$5)\sim R_{{\mathfrak h}^0_{1.3}}$,
$6)\sim R_{{\mathfrak h}_{1.1}}$,
$7)\sim R_{{\mathfrak h}^{x_3}_{2.4}}$,
$8)\sim R_{{\mathfrak h}^0_{2.4}}$.

\subsection{The reduced extended Galilei algebra $\boldsymbol{{\rm A\bar G}_2(1)}$}\label{sec_AG_21}

The reduced extended Galilei algebra ${\rm A \bar G}_2(1)$ of a one-dimensional space is generated by
four operators $P$, $G$, $D$ and $T$ that are connected by commutation
relations $[G,T]=P$, $[D,G]=G$, $[P,D]=P$ and $[T,D]=2T$.

${\rm A\bar G}_2(1)$ is isomorphic to a Lie algebra $A^{b=-1/2}_{4.8}$ with the commutators
$[e_2,e_3]=e_1$,
$[e_1,e_4]=\tfrac 12e_1$,
$[e_2,e_4]=e_2$ and
$[e_3,e_4]=-\tfrac 12e_3$, where the isomorphism is established by the equalities{\samepage
\begin{equation}\label{baseAG21}
P=-e_1,\quad T=e_2,\quad G=e_3,\quad D=2e_4.
\end{equation}
Faithful realizations of $A^{-1/2}_{4.8}$ are given in Table~\ref{tab:faithfreala1248}.}

\begin{table*}\small
\renewcommand{\arraystretch}{1.1}
\centering
\begin{tabular}{lrlllll}
\hline
&1) & $e_1=\p_1$, & $e_2=\p_2$,&  $e_3=x_2\p_1+\p_3$,& $e_4=\tfrac 12 x_1\p_1+x_2\p_2-\tfrac 12 x_3\p_3+\p_4$ & \\
&2) & $e_1=\p_1$, & $e_2=\p_2$, & $e_3=x_2\p_1+\p_3$,& $e_4=\tfrac 12 x_1\p_1+x_2\p_2-\tfrac 12 x_3\p_3$&\\
&3) & $e_1=\p_1$, & $e_2=\p_2$, & $e_3=x_2\p_1+x_3\p_2$, & $e_4=\tfrac 12 x_1\p_1+x_2\p_2+\tfrac 32 x_3\p_3$  & \\
&4) & $e_1=\p_1$, & $e_2=\p_2$, & $e_3=x_2\p_1$, & $e_4=\tfrac 12 x_1\p_1+x_2\p_2+\p_3$ &\\
&5) & $e_1=\p_1$, & $e_2=\p_2$, & $e_3=x_2\p_1$, & $e_4=\tfrac 12 x_1\p_1+x_2\p_2$ &\\
&6) & $e_1=\p_1$, & $e_2=x_2\p_1$, & $e_3=-\p_2$, &  $e_4=\tfrac 12 x_1\p_1-\tfrac 12x_2\p_2+\p_3$ &\\
&7) & $e_1=\p_1$, & $e_2=x_2\p_1$, & $e_3=-\p_2$, & $e_4=\tfrac 12 x_1\p_1-\tfrac 12x_2\p_2$&\\
\hline
\end{tabular}
%\end{ruledtabular}
\caption{Faithful realizations of $A^{-1/2}_{4.8}$.}
\label{tab:faithfreala1248}
\end{table*}

The full group of automorphisms of ${\rm A\bar G}_2(1)=\langle P,\, T,\, G,\, D\rangle$ is generated by the nonsingular matrices
\begin{equation*}
\begin{pmatrix}
\alpha_{22}\alpha_{33}& 0 & 0& 0 \\
-\alpha_{22}\alpha_{43}& \alpha_{22} &0&0\\
-\frac12\alpha_{42}\alpha_{33}&0&\alpha_{33}&0\\
\alpha_{41}& \alpha_{42} &\alpha_{43} & 1
\end{pmatrix},\quad \mbox{where}\ \alpha_{ij} \in \R,\ \mbox{and}\  \alpha_{22}\alpha_{33}\neq0.
\end{equation*}

It follows from $\Aut({\rm A\bar G}_2(1))$ and results of~\cite{Fushchych&Barannyk&Barannyk} that the list of inequivalent subalgebras is as follows
\begin{alignat*}{4}
&\mathfrak h_{1.1}= \langle P \rangle, \quad & & \mathfrak h_{2.1}= \langle P,\,T \rangle, \quad &&\mathfrak h_{3.1}= \langle P,\,T,\,G \rangle,&\\
&\mathfrak h_{1.2}= \langle T \rangle, \quad & & \mathfrak h_{2.2}= \langle P,\,G \rangle, \quad && \mathfrak h_{3.2}= \langle P,\,T,\,D \rangle,&\\
&\mathfrak h_{1.3}= \langle G \rangle, \quad && \mathfrak h_{2.3}= \langle P,\,D \rangle, \quad && \mathfrak h_{3.3}= \langle P,\,G,\,D \rangle.&\\
&\mathfrak h_{1.4}= \langle  D \rangle, \quad && \mathfrak h_{2.4}= \langle P, \,T+G \rangle,\quad &&&\\
&\mathfrak h_{1.5}= \langle T+G \rangle; \quad && \mathfrak h_{2.5}= \langle T,\,D \rangle,\quad &&&\\
&&& \mathfrak h_{2.6}= \langle G, \,D \rangle;\quad &&&
\end{alignat*}

The ideals of ${\rm A\bar G}_2(1)$ are
$\langle P \rangle$,
$\langle P, \,T\rangle$,
$\langle P, \,G \rangle$
and
$\langle P, \,T, \,G\rangle$, therefore the realizations related to
$\mathfrak h_{1.1}$, $\mathfrak h_{2.1}$, $\mathfrak h_{2.2}$, $\mathfrak h_{2.3}$, $\mathfrak h_{2.4}$,
$\mathfrak h_{3.1}$, $\mathfrak h_{3.2}$ and $\mathfrak h_{3.3}$ should be unfaithful.

The list of realizations is presented in Table~\ref{tab:a_g21a}.

\begin{table*}\small
\renewcommand{\arraystretch}{1.1}\centering
\begin{tabular}{rl}
\hline
&\\[-8pt]
\parbox{2.7 cm}{Complementary\\ basis} & Realization\\[8pt]
\hline
&\\[-8pt]
$\{P,\,T,\,G,\,D\}$&$R_{\mathfrak h_{0}}\colon\
P=\partial_1,\
T=\partial_2,\
G=-x_2 \partial_1+\partial_3,\
D=x_1 \partial_1+2 x_2 \partial _2-x_3 \partial _3+\partial _4$
\\
% subalgebra {P}
$\{T,\,G,\,D\}$&$  R_{\mathfrak h_{1.1}}\colon\
P=0,\
T=\partial _1,\
G=\partial _2,\
D=2 x_1 \partial _1-x_2 \partial _2+\partial _3$
\\
% subalgebra {T}
$\{P,\,G,\,D\}$&$  R_{\mathfrak h_{1.2}}\colon\
P=\partial _1,\
T=x_2 \partial _1,\
G=\partial _2,\
D=x_1 \partial _1-x_2 \partial _2+\partial _3$
\\
% subalgebra {G}
$\{P,\,T,\,D\}$&$  R_{\mathfrak h_{1.3}}\colon\
P=\partial _1,\
T=\partial _2,\
G=-x_2 \partial _1,\
D=x_1 \partial _1+2 x_2 \partial _2+\partial _3$
\\
% subalgebra {D}
$\{P,\,T,\,G\}$&$  R_{\mathfrak h_{1.4}}\colon\
P=\partial _1,\
T=\partial _2,\
G=-x_2 \partial_1+\partial_3,\
D=x_1 \partial _1+2 x_2 \partial _2-x_3 \partial _3$
\\
% subalgebra {G+T}
$\{P,\,T,\,D\}$&$  R_{\mathfrak h_{1.5}}\colon\
P=\partial _1,\
T=\partial _2,\
G=-x_2 \partial _1-e^{3 x_3} \partial _2,\
D=x_1 \partial _1+2 x_2 \partial _2+\partial _3$
\\
% subalgebra {P,T}
$\{G,\,D\}$&$ R_{\mathfrak h_{2.1}}\colon\
P=0,\
T=0,\
G=\partial _1,\
D=-x_1 \partial _1+\partial _2$
\\
% subalgebra {P,G}
$\{T,\,D\}$&$ R_{\mathfrak h_{2.2}}\colon\
P=0,\
T=\partial _1,\
G=0,\
D=2 x_1 \partial _1+\partial _2$
\\
% subalgebra {P,D}
$\{T,\,G\}$&$  R_{\mathfrak h_{2.3}}\colon\
P=0,\
T=\partial _1,\
G=\partial _2,\
D=2 x_1 \partial _1-x_2 \partial _2$
\\
% subalgebra {P,G+T}
$\{T,\,D\}$&$  R_{\mathfrak h_{2.4}}\colon\
P=0,\
T=\partial _1,\
G=-e^{3 x_2} \partial _1,\
D=2 x_1 \partial _1+\partial _2$
\\
% subalgebra {T,D}
$\{P,\,G\}$&$  R_{\mathfrak h_{2.5}}\colon\
P=\partial _1,\
T=x_2 \partial _1,\
G=\partial _2,\
D=x_1 \partial _1-x_2 \partial _2$
\\
% subalgebra {G,D}
$\{P,\,T\}$&$  R_{\mathfrak h_{2.6}}\colon\
P=\partial _1,\
T=\partial _2,\
G=-x_2 \partial _1,\
D=x_1 \partial _1+2 x_2 \partial _2$
\\
% subalgebra {P,T,G}
$\{D\}$&$  R_{\mathfrak h_{3.1}}\colon\
P=0,\
T=0,\
G=0,\
D=\partial _1$
\\
% subalgebra {P,T,D}
$\{G\}$&$  R_{\mathfrak h_{3.2}}\colon\
P=0,\
T=0,\
G=\partial _1,\
D=-x_1 \partial _1$
\\
% subalgebra {P,G,D}
$\{T\}$ &$R_{\mathfrak h_{3.3}}\colon\
P=0,\
T=\partial _1,\
G=0,\
D=2 x_1 \partial _1$
\\
\hline
\end{tabular}
%\end{ruledtabular}
\caption{Realizations of the reduced extended Galilei algebra ${{\rm A\bar{G}}_2(1)}$.}
\label{tab:a_g21a}
\end{table*}

All constructed faithful realizations of ${\rm A\bar G}_2(1)$ are equivalent to those of $A^{-1/2}_{4.8}$ listed in~\cite{Popovych2003}. Namely, the following cases are equivalent:
$1)\sim R_{{\mathfrak h}_0}$,
$2)\sim R_{{\mathfrak h}_{1.4}}$,
$3)\sim R_{{\mathfrak h}_{1.5}}$,
$4)\sim R_{{\mathfrak h}_{1.3}}$,
$5)\sim R_{{\mathfrak h}_{2.6}}$,
$6)\sim R_{{\mathfrak h}_{1.2}}$,
$7)\sim R_{{\mathfrak h}_{2.5}}$.

To establish connection $R_{{\mathfrak h}_{1.5}}\mapsto 3)$, the change of variables $x_2\mapsto-x_2$, $-e^{3x_3}\mapsto x_3$ should be performed.
Transformations in the rest of cases are quite obvious and exhausted by those of alternating signs and scalings of variables.

\section{Five-dimensional cases}\label{sec_5d} There are two five-dimensional Galilei algebras: ${\rm AG}_2(1)=\langle M,\,P,\,T,\,G,\,D\rangle$ with commutation relations
$[T,G]=-P,$ $[D,G]=G$, $[P,D]=P$,  $[T,D]=2T$, $[G,P]=M$ and
${\rm A\bar{G}}_3(1)=\langle P,\,T,\,G,\,D,\,S\rangle$ with commutation relations
$[T,G]=-P,$ $[D,G]=G$, $[P,D]=P$,  $[T,D]=2T$, $[S,P]=G$, $[T,S]=D$, $[D,S]=2S$.

\subsection{The extended Galilei algebra $\boldsymbol{{\rm AG}_2(1)}$}\label{sec_AG21}
The Galilei algebra ${\rm AG}_2(1)=\langle M,\,P,\,T,\,G,\,D\rangle$ is isomorphic to the solvable Lie algebra $\mathfrak{g}^{-2}_{5.30}$ in the classification by Mubarakzyanov~\cite{mubarakzyanov1963.2}.
The change of the basis is given by the formulae
\begin{equation}\label{baseAG21e}
M=e_1,\quad  P=-e_2,\quad  T=e_3,\quad  G=e_4,\quad  D=-e_5.
\end{equation}

The whole automorphism group $\Aut({\rm AG}_2(1))$ is generated by matrices of the form
\begin{equation*}
\begin{pmatrix}
 \alpha_{22} \alpha_{44} & 0 & 0 & 0 & 0 \\
 -\alpha_{22} \alpha_{54} & \alpha_{22} & 0 & 0 & 0 \\
 \frac{\alpha_{22} \alpha_{54}^2}{2 \alpha_{44}} & -\frac{\alpha_{22} \alpha_{54}}{\alpha_{44}} & \frac{\alpha_{22}}{\alpha_{44}} & 0
   & 0 \\
 \alpha_{42} \alpha_{54}-\alpha_{44} \alpha_{52} & \alpha_{42} & 0 & \alpha_{44} & 0 \\
 \alpha_{51} & \alpha_{52} & -\frac{2 \alpha_{42}}{\alpha_{44}} & \alpha_{54} & 1 \\
\end{pmatrix},
\end{equation*}
where $\alpha_{22},\alpha_{44} \in \mathbb{R}{\setminus} \{0\}$ and other parameters are arbitrary.

All proper subalgebras of ${\rm AG}_2(1)$ inequivalent with respect to inner automorphisms are
\begin{alignat*}{4}
&\mathfrak h_{1.1}=\langle M\rangle, \quad && \mathfrak h_{2.1}=\langle M, \,D\rangle,\quad &&   \mathfrak h_{3.1}=\langle M, \,T, \,D\rangle, &\\
&\mathfrak h_{1.2}=\langle D\rangle, \quad &&  \mathfrak h_{2.2}=\langle M, \,T\rangle,\quad && \mathfrak h_{3.2}=\langle M, \,G, \,D\rangle,& \\
&\mathfrak h_{1.3}=\langle T\rangle, \quad && \mathfrak h_{2.3}=\langle M, \,G\rangle, \quad &&  \mathfrak h_{3.3}=\langle M, \,P, \,D\rangle, & \\
&\mathfrak h_{1.4}=\langle G\rangle, \quad && \mathfrak h_{2.4}=\langle M,\, P\rangle, \quad && \mathfrak h_{3.4}=\langle M, \,P,\, T\rangle, & \\
&\mathfrak h_{1.5}=\langle P\rangle, \quad && \mathfrak h_{2.5}=\langle T, \,D\rangle, \quad && \mathfrak h_{3.5}=\langle M, \,P,\, G\rangle,& \\
&\mathfrak h_{1.6}=\langle M+T\rangle, \quad && \mathfrak h_{2.6}=\langle G, \,D\rangle, \quad && \mathfrak h_{3.6}=\langle P, \,T,\, D\rangle,& \\  &\mathfrak h_{1.7}=\langle G+T\rangle, \quad &&\mathfrak h_{2.7}=\langle P,\, D\rangle, \quad &&  \mathfrak h_{3.7}=\langle M,\, P,\, G+T\rangle,& \\
&\mathfrak h_{1.8}=\langle D+\alpha  M\rangle, \quad && \mathfrak h_{2.8}=\langle P,\, T\rangle, \quad && \mathfrak h_{3.8}=\langle P, \,T, \,D+\alpha  M\rangle;&\\
&\mathfrak h_{1.9}=\langle T-M\rangle; \quad && \mathfrak h_{2.9}=\langle P,\, M+T\rangle,\quad && \mathfrak h_{4.1}=\langle M, \,P, \,T, \, D\rangle,&\\
&&& \mathfrak h_{2.10}=\langle M, \, G+T\rangle,\quad && \mathfrak h_{4.2}=\langle M, \,P, \,G,\, D\rangle, & \\
&&& \mathfrak h_{2.11}=\langle T,\, D+\alpha  M\rangle,\quad && \mathfrak h_{4.3}=\langle M, \,P, \,T, \,G\rangle,& \\
&&& \mathfrak h_{2.12}=\langle G, \, D+\alpha  M\rangle,\quad &&&\\
&&& \mathfrak h_{2.13}=\langle P, \, D+\alpha  M\rangle,\quad &&&\\
&&& \mathfrak h_{2.14}=\langle P, \, T-M\rangle; \quad &&&
\end{alignat*}
where $\alpha \neq 0$.
The algebras $\mathfrak h_{1.1}$ and ${\mathfrak h}^\alpha_{1.8}$ for $\alpha \neq 0$ are isomorphic via $\left(\begin{smallmatrix}
 \frac{1}{\alpha } & 0 & 0 & 0 & 0 \\
 0 & 1 & 0 & 0 & 0 \\
 0 & 0 & \alpha  & 0 & 0 \\
 0 & 0 & 0 & \frac{1}{\alpha } & 0 \\
 -\frac{1}{\alpha } & 0 & 0 & 0 & 1 \\
\end{smallmatrix}\right)$. All other subalgebras are nonisomorphic with respect to the whole automorphism group. They correspond to the realizations given in Table~\ref{tab:ag21c}.

\begin{table*}[b!]\small
\renewcommand{\arraystretch}{1.1}\centering
\begin{tabular}{rl}
\hline
&\\[-8pt]
\parbox{2.7 cm}{Complementary\\ basis} & Realization\\[8pt]
\hline
& \\[-8pt]
$\{M,\,P,\,T,\,D,\,G\}$& $R_{\mathfrak h_0}\colon\
M=\partial _1,\
P=\partial _2,\
T=\partial _3,\
G=-x_2 \partial _1-x_3 \partial_2+e^{x_4} \partial_5,$
\\
& $\phantom{R_{\mathfrak h_0}\colon\ }D=x_2 \partial_2+2 x_3 \partial_3+\partial_4$
\\
% subalgebra {M}
$\{P,\,T,\,D,\,G\}$& $R_{\mathfrak h_{1.1}}\colon\
M=0,\
P=\partial _1,\
T=\partial _2,\
G=-x_2 \partial _1+e^{x_3} \partial _4,\ D=x_1 \partial_1+2 x_2 \partial_2+\partial_3$
\\
$\{M,\,P,\,T,\,G\}$ & $R_{\mathfrak h_{1.2}}\colon\
M=\partial _1,\
P=\partial _2,\
T=\partial _3,\
G=-x_2 \partial _1-x_3 \partial _2+\partial _4,$
\\ &
$\phantom{R_{\mathfrak h_{1.1}}\colon\ }D=x_2 \partial_2+2 x_3 \partial_3-x_4 \partial_4$ \\
% subalgebra {T}
$\{M,\,G,\,P,\,D\}$& $R_{\mathfrak h_{1.3}}\colon\
M=\partial _1,\
P=x_2 \partial _1+\partial _3,\
T=\frac{1}{2} x_2^2 \partial _1+x_2 \partial _3,\
G=\partial _2,$
\\
& $\phantom{R_{\mathfrak h_{1.1}}\colon\ }D=-x_2 \partial_2+x_3 \partial_3+\partial_4$
\\
% subalgebra {G}
$\{M,\,P,\,T,\,D\}$  & $R_{\mathfrak h_{1.4}}\colon\
M=\partial _1,\
P=\partial _2,\
T=\partial _3,\
G=-x_2 \partial _1-x_3 \partial _2,\ D=x_2 \partial _2+2 x_3 \partial _3+\partial _4$\\
% subalgebra {P}
$\{M,\,T,\,G,\,D\}$& $R_{\mathfrak h_{1.5}}\colon\
M=\partial _1,\
P=x_3 \partial _1,\
T=\partial _2,\
G=-x_2 x_3 \partial _1+\partial _3,\ D=2 x_2 \partial _2-x_3 \partial _3+\partial _4$ \\
% subalgebra {M+T}
$\{P,\,G,\,D,\,T\}$& $R_{\mathfrak h_{1.6}}\colon\
M=-\partial _4,\
P=\partial _1,\
T=x_2 \partial _1+\left(\frac1{2}{x_2^2}+e^{-2 x_3}\right) \partial _4,\ G=\partial _2+x_1 \partial _4,$
\\ & $\phantom{R_{\mathfrak h_{1.1}}\colon\ }
D=x_1 \partial _1-x_2 \partial _2+\partial _3$\\
% subalgebra {G+T}
$\{M,P,T,D\}$& $R_{\mathfrak h_{1.7}}\colon\
M=\partial _1,\
P=\partial _2,\
T=\partial _3,\
G=-x_2 \partial _1-x_3 \partial _2-e^{3 x_4} \partial _3,$
\\
& $\phantom{R_{\mathfrak h_{1.1}}\colon\ }D=x_2 \partial _2+2 x_3 \partial _3+\partial _4$
\\
$\{G,\,P,\,D,\,T\}$ & $R_{\mathfrak h_{1.9}}\colon\
M=\partial _4,\
P=\partial _2+x_1 \partial _4,\
T=x_1 \partial _2+\left(\frac1{2}{x_1^2}+e^{-2 x_3}\right) \partial_4$
\\ & $\phantom{R(\mathfrak h_{1.1})\colon\ }G=\partial _1,\
D=-x_1 \partial _1+x_2 \partial_2+\partial _3$\\
% subalgebra {M,D}
$\{P,\,G,\,T\}$& $R_{\mathfrak h_{2.1}}\colon\
M=0,\
P=\partial _1,\
T=x_2 \partial _1+\partial _3,\
G=\partial _2,\ D=x_1 \partial_1-x_2 \partial _2+2 x_3 \partial _3$ \\
% subalgebra {M,T}
$\{P,\,G,\,D\}$& $R_{\mathfrak h_{2.2}}\colon\
M=0,\
P=\partial _1,\
T=x_2 \partial _1,\
G=\partial_2,\ D=x_1 \partial_1-x_2 \partial_2+\partial_3$
\\
% subalgebra {M,G}
$\{P,\,T,\,D\}$& $R_{\mathfrak h_{2.3}}\colon\
M=0,\
P=\partial _1,\
T=\partial _2,\
G=-x_2 \partial_1,\
D=x_1 \partial_1+2 x_2 \partial_2+\partial_3$
\\
% subalgebra {M,P}
$\{D,\,T,\,G\}$& $R_{\mathfrak h_{2.4}}\colon\
M=0,\
P=0,\
T=e^{-2 x_1} \partial _2,\
G=e^{x_1} \partial _3,\
D=\partial _1$
\\
% subalgebra {T,D}
$\{M,\,G,\,P\}$& $R_{\mathfrak h_{2.5}}\colon\
M=\partial _1,\
P=x_2 \partial _1+\partial _3,\
T=\frac{1}{2} x_2^2 \partial _1+x_2 \partial _3,\
G=\partial _2,\ D=-x_2 \partial _2+x_3 \partial _3$ \\
% subalgebra {G,D}
$\{M,\,P,\,T\}$& $R_{\mathfrak h_{2.6}}\colon\
M=\partial _1,\
P=\partial _2,\
T=\partial _3,\
G=-x_2 \partial _1-x_3 \partial _2,\
D=x_2 \partial _2+2 x_3 \partial _3$
\\
% subalgebra {P,D}
$\{M,\,T,\,G\}$& $R_{\mathfrak h_{2.7}}\colon\
M=\partial _1,\
P=x_3 \partial _1,\
T=\partial _2,\
G=-x_2 x_3 \partial _1+\partial _3,\
D=2 x_2 \partial _2-x_3 \partial_3$\\
% subalgebra {P,T}
$\{M,G,D\}$& $R_{\mathfrak h_{2.8}}\colon\
M=\partial _1,\
P=x_2 \partial _1,\
T=\frac{1}{2} x_2^2 \partial _1,\
G=\partial _2,\
D=-x_2 \partial _2+\partial _3$
 \\
% subalgebra {P,M+T}
$\{G,\,D,\,T\}$& $R_{\mathfrak h_{2.9}}\colon\
M=-\partial _3,\
P=-x_1 \partial _3,\
T=\left(-\frac1{2}{x_1^2}+e^{-2 x_2}\right) \partial _3,\
G=\partial _1,\
D=-x_1 \partial _1+\partial _2$ \\
% subalgebra {M,G+T}
$\{P,\,T,\,D\}$& $R_{\mathfrak h_{2.10}}\colon\
M=0,\
P=\partial _1,\
T=\partial _2,\
G=-x_2 \partial _1-e^{3 x_3} \partial _2,\ D=x_1 \partial _1+2 x_2 \partial _2+\partial _3$\\
$\{P,G,D\}$& $R_{\mathfrak h_{2.11}}\colon\
M=-\frac1{\alpha}{\partial_3},\
P=\partial _1,\
T=x_2 \partial _1+\frac1{2 \alpha }{x_2^2 \partial _3},\
G=\partial _2+\frac1{\alpha }{x_1 \partial _3},$
\\
& $\phantom{R(\mathfrak h_{2.11})\colon\ }D=x_1 \partial_1-x_2 \partial_2+\partial_3$
\\
% subalgebra {G,D+M \[Alpha]}
$\{P,\,T,\,D\}$& $R_{\mathfrak h_{2.12}}\colon\
M=-\frac1{\alpha}{\partial_3},\
P=\partial _1,\
T=\partial _2,\
G=-x_2 \partial _1+\frac1{\alpha }{x_1 \partial _3},\
D=x_1 \partial _1+2 x_2 \partial _2+\partial _3$
\\
% subalgebra {P,D+M \[Alpha]}
$\{T,\,G,\,D\}$& $R_{\mathfrak h_{2.13}}\colon\
M=-\frac1{\alpha}{\partial_3},\
P=-\frac1{\alpha}{x_2 \partial _3},\
T=\partial _1,\
G=\partial _2+\frac1{\alpha}{x_1 x_2 \partial _3},$\\
&$
\phantom{R(\mathfrak h_{2.11})\colon\ } D=2 x_1 \partial _1-x_2 \partial_2+\partial_3$
\\
% subalgebra {P,-M+T}
$\{G,\,D,\,T\}$& $R_{\mathfrak h_{2.14}}\colon\
M=\partial _3,\
P=x_1 \partial _3,\
T=\left(\frac1{2}{x_1^2}+e^{-2 x_2}\right) \partial _3,\
G=\partial _1,\ D=-x_1 \partial _1+\partial _2$
\\
% subalgebra {T,D+M \[Alpha]}
% subalgebra {M,T,D}
$\{P,\,G\}$& $R_{\mathfrak h_{3.1}}\colon\
M=0,\
P=\partial_1,\
T=x_2 \partial_1,\
G=\partial_2,\
D=x_1 \partial_1-x_2 \partial_2$\\
% subalgebra {M,G,D}
$\{P,\,T\}$& $R_{\mathfrak h_{3.2}}\colon\
M=0,\
P=\partial _1,\
T=\partial _2,\
G=-x_2 \partial _1,\
D=x_1 \partial _1+2 x_2 \partial _2$ \\
% subalgebra {M,P,D}
$\{T,\,G\}$& $R_{\mathfrak h_{3.3}}\colon\
M=0,\
P=0,\
T=\partial _1,\
G=\partial _2,\
D=2 x_1 \partial _1-x_2 \partial _2$ \\
% subalgebra {M,P,T}
$\{D,\,G\}$& $R_{\mathfrak h_{3.4}}\colon\
M=0,\
P=0,\
T=0,\
G=e^{x_1} \partial _2,\
D=\partial _1$
\\
% subalgebra {M,P,G}
$\{D,\,T\}$& $R_{\mathfrak h_{3.5}}\colon\
M=0,\
P=0,\
T=e^{-2 x_1} \partial _2,\
G=0,\
D=\partial_1$
\\
% subalgebra {P,T,D}
$\{M,\,G\}$& $R_{\mathfrak h_{3.6}}\colon\
M=\partial _1,\
P=x_2 \partial _1,\
T=\frac{1}{2} x_2^2 \partial _1,\
G=\partial _2,\
D=-x_2 \partial _2$
\\
% subalgebra {M,P,G+T}
$\{D,\,T\}$& $R_{\mathfrak h_{3.7}}\colon\
M=0,\
P=0,\
T=e^{-2 x_1} \partial _2,\
G=-e^{x_1} \partial _2,\
D=\partial _1$
\\
% subalgebra {P,T,D+M \[Alpha]}
$\{G,\,D\}$& $R_{\mathfrak h_{3.8}}\colon\
M=-\frac1{\alpha }{\partial _2},\
P=-\frac1{\alpha }{x_1 \partial _2},\
T=-\frac1{2 \alpha}{x_1^2 \partial _2},\
G=\partial _1,\
D=-x_1 \partial _1+\partial _2$ \\
% subalgebra {M,P,T,D}
$\{G\}$& $R_{\mathfrak h_{4.1}}\colon\
M=0,\
P=0,\
T=0,\
G=\partial _1,\
D=-x_1 \partial _1$
\\
% subalgebra {M,P,G,D}
$\{T\}$& $R_{\mathfrak h_{4.2}}\colon\
M=0,\
P=0,\
T=\partial_1,\
G=0,\
D=2 x_1 \partial_1$
\\
% subalgebra {M,P,T,G}
$\{D\}$& $R_{\mathfrak h_{4.3}}\colon\
M=0,\
P=0,\
T=0,\
G=0,\
D=\partial_1$
\\
\hline
\end{tabular}
\caption{Realizations of the extended Galilei algebra ${{\rm A{G}}_2(1)}$.}\label{tab:ag21c}
\end{table*}

\subsection{The reduced special Galilei algebra $\boldsymbol{{\rm A\bar{G}}_3(1)}$}\label{sec_AG_31}
The reduced special Galilei algebra ${\rm A\bar{G}}_3(1)=\langle P,\,T,\,G,\,D,\,S\rangle$ is isomorphic to the nonsolvable and nondecomposable Lie algebra $\mathfrak{g}_{5}$ ($\mathfrak{sl}(2,\mathbb R)\lsemioplus 2A_1$) in the classification by Mubarakzyanov~\cite{mubarakzyanov1963.2}.
It should be mentioned that there is a~misprint in~\cite{mubarakzyanov1963.2}, namely, the adduced commutation relation $[e_2,e_3]=-2e_3$ should be $[e_2,e_3]=2e_3$, note, that this Mubarakzyanov's classification was already enhanced in~\cite{Patera}, see also tables in arXiv version of~\cite{BoykoPateraPopovych2006}.
The change of the basis is given by the formulae
\begin{equation*}
T=e_1,\quad D=e_2,\quad S=e_3,\quad G=e_4,\quad P=e_5.
\end{equation*}

Up to inner automorphism, the Galilei algebra ${\rm A\bar{G}}_3(1)$ has the following proper subalgebras:
\begin{alignat*}{4}
&\mathfrak h_{1.1}=\langle D\rangle, \quad && \mathfrak h_{2.1}=\langle D, \,T\rangle,  \quad & & \mathfrak h_{3.1}=\langle D,\, S,\, T\rangle,& \\
&\mathfrak h_{1.2}=\langle T\rangle,  \quad  & & \mathfrak h_{2.2}=\langle D, \,P\rangle, \quad & &\mathfrak h_{3.2}=\langle D,\, T, \,P\rangle,& \\
&\mathfrak h_{1.3}=\langle P\rangle,   \quad  & & \mathfrak h_{2.3}=\langle T,\, P\rangle, \quad & &\mathfrak h_{3.3}=\langle D, \,G,\, P\rangle,& \\
&\mathfrak h_{1.4}=\langle S+T\rangle,   \quad  & & \mathfrak h_{2.4}=\langle G, \,P\rangle,\quad && \mathfrak h_{3.4}=\langle T,\, G,\, P\rangle,& \\
&\mathfrak h_{1.5}=\langle G+T\rangle,  \quad & &  \mathfrak h_{2.5}=\langle G+T, \,P\rangle,\quad && \mathfrak h_{3.5}=\langle S+T,\, G,\, P\rangle; &\\
&\mathfrak h_{1.6}=\langle T-G\rangle; \quad && \mathfrak h_{2.6}=\langle T-G, \, P\rangle;\quad && \mathfrak h_{4.1}=\langle D, \,T, \,G, \,P\rangle. \end{alignat*}
As the subalgebras $\mathfrak h_{1.5}$ and $\mathfrak h_{1.6}$ are isomorphic via outer automorphism
$\left(\begin{smallmatrix}
 1 & 0 & 0 & 0 & 0 \\
 0 & -1 & 0 & 0 & 0 \\
 0 & 0 & -1 & 0 & 0 \\
 0 & 0 & 0 & 1 & 0 \\
 0 & 0 & 0 & 0 & 1 \\
\end{smallmatrix}\right)$, we exclude $\mathfrak h_{1.6}$ from the list.
Then the above subalgebras are pairwise nonisomorphic with respect to the whole automorphism group, which is generated
by these five nonsingular matrices with arbitrary parameters $\alpha _{ij}\in \R$ and
corresponding realizations are given in Table~\ref{tab:algag31}.
\begin{align*}
%%%%%%%%%%% case 0 %%%%%%%%%%%%%%%%%%%%%%%%%%%%%
&\left(\begin{smallmatrix}
\frac {\alpha_{11}\alpha_{22}(2+\alpha_{11})}{2\alpha_{14}\alpha_{41}}&0&\frac {-\alpha_{11}\alpha_{22}}{2\alpha_{41}}&0&0\\
\frac{\alpha_{11}(2+\alpha_{11})(\alpha_{12}(2+\alpha_{11})-2\alpha_{13})}{-4\alpha_{14}\alpha_{41}}&
\frac {\alpha_{11}(2+\alpha_{11}) ^{2}}{4\alpha_{14}\alpha_{41}}&
\frac {\alpha_{12} (2+\alpha_{11}) -2\alpha_{13}}{4\alpha_{41}}&
\frac {\alpha_{11} ( 2+\alpha_{11})}{4\alpha_{41}}&
\frac {\alpha_{14}\alpha_{11}}{4\alpha_{41}}\\
\frac {-\alpha_{11}\alpha_{22}}{\alpha_{14}}&0&\alpha_{22}&0&0\\
\frac{\alpha_{13}}{\alpha_{14}}&
{\frac {\alpha_{11} (2+\alpha_{11})}{\alpha_{14}}}&
\frac{\alpha_{12}-\alpha_{13}}{\alpha_{11}}&
1+\alpha_{11}&
\alpha_{14}\\
\frac {\alpha_{41}\alpha_{12}}{\alpha_{14}}&
\frac {\alpha_{41}\alpha_{11}}{\alpha_{14}}&
\frac {-\alpha_{41}\alpha_{12}}{\alpha_{11}}&
\alpha_{41}&
\frac {\alpha_{14}\alpha_{41}}{\alpha_{11}}
\end{smallmatrix}\right),\ \alpha_{11}\alpha_{22}\alpha_{14}\alpha_{41}\ne0;\\[1ex]
%%%%%%%%%% case 1 %%%%%%%%%%%%%%%%%%%%%%%%%%%%%
&\left(\begin{smallmatrix}
a_{23}a_{51}&0&\frac{2a_{23}a_{51}}{a_{15}}&0&0\\
\frac12 a_{51} a_{13}&-\frac 12 a_{15}a_{51}&\frac {a_{51}a_{13}}{a_{15}}&a_{51}& \frac{-2a_{51}}{a_{15}}\\
a_{23}&0&0&0&0\\
\frac12(a_{12}a_{15}-a_{13})&a_{15}&a_{12}&-1&0\\
\frac{-a_{12}a_{15}}{2a_{51}}&-\frac{a_{15}}{2a_{51}}&0&0&0
\end{smallmatrix}\right),\quad \text{where }\alpha_{23}\alpha_{51}\alpha_{15}\ne0;\\[1ex]
%%%%%%%%%%% case 2 %%%%%%%%%%%%%%%%%%%%%%%%%%%%%
&\left(\begin{smallmatrix}
\frac{2\alpha_{22}\alpha_{51}}{\alpha_{14}}&0&-\alpha_{22}\alpha_{51}&0&0\\
-\frac{\alpha_{51}\alpha_{12}}{\alpha_{14}}& \frac {2\alpha_{51}}{\alpha_{14}}&\frac 12 \alpha_{51}\alpha_{12}&\alpha_{51}&\frac 12\alpha_{14}\alpha_{51}\\
0&0&\alpha_{22}&0&0\\
\alpha_{13}&0&\frac 12(\alpha_{12}-\alpha_{13}\alpha_{14})&1&\alpha_{14}\\
0&0&\frac {-\alpha_{13}\alpha_{14}}{2\alpha_{51}}&0&\frac {\alpha_{14}}{2\alpha_{51}}
\end{smallmatrix}\right),\quad \text{where }\alpha_{14}\alpha_{22}\alpha_{51}\ne0;\\[1ex]
%%%%%%%%%%% case 3 %%%%%%%%%%%%%%%%%%%%%%%%%%%%%
&\left(\begin{smallmatrix}
\alpha_{22}\alpha_{44}&0&0&0&0\\
-\alpha_{12}\alpha_{44} & \alpha_{44}&0&0&0\\
-\alpha_{22} \alpha_{41}\alpha_{44}& 0 & \alpha_{22}&0&0\\
(\alpha_{13}-\alpha_{12}\alpha_{41})\alpha_{44}&2\alpha_{41}\alpha_{44}& \alpha_{12}&1&0\\
\alpha_{41}\alpha_{13}\alpha_{44}&\alpha_{41}^2\alpha_{44}&-\alpha_{13}& \alpha_{41}& \frac{1}{\alpha_{44}}
\end{smallmatrix}\right),\ \alpha_{22}\alpha_{44}\ne0;\quad
%%%%%%%%%%% case 4 %%%%%%%%%%%%%%%%%%%%%%%%%%%%%
\left(\begin{smallmatrix}
0&0&-\alpha_{23}\alpha_{54}&0&0\\
0&0&\alpha_{13}\alpha_{54}&0&\alpha_{54}\\
\alpha_{23}&0&0&0&0\\
\alpha_{13}&0&\alpha_{12}&-1&0\\
\frac{\alpha_{12}}{\alpha_{54}}&\frac{1}{\alpha_{54}}&0&0&0\\
\end{smallmatrix}\right),\ \alpha_{23}\alpha_{54}\ne0.
\end{align*}

\begin{remark}
The results of this section are also valuable for abstract theory of Lie algebras as far as it gives the complete classification of inequivalent realizations of two real five-dimensional Lie algebras $\mathfrak{g}^{-2}_{5.30}$ and $\mathfrak{g}_{5}$~\cite{mubarakzyanov1963.2}, which were not known before.
\end{remark}

\begin{table*}[t!]\small
\renewcommand{\arraystretch}{1.1}
\centering
\begin{tabular}{rl}
\hline
&\\[-8pt]
\parbox{2.7 cm}{Complementary\\ basis} & Realization\\[8pt]
\hline
&\\[-8pt]
$\{G,\,P,\,T,\,D,\,S\}$&
$R_{\mathfrak h_{0}}\colon\
D=-x_1 \partial_1+x_2 \partial_2+2 x_3 \partial_3+\partial_4,\
G=\partial_1,\
P=\partial_2,$
\\
&$\phantom{R_{\mathfrak h_{0}}\colon\ }S=-x_2 \partial_1+x_3^2 \partial _3+x_3 \partial_4+e^{2 x_4} \partial_5,\
T=x_1 \partial_2+\partial_3$
\\
$\{G,\,P,\,S,\,T\}$ &$ R_{\mathfrak h_{1.1}}\colon\
D=-x_1\partial_1+x_2 \partial_2-2 x_3 \partial_3+2 x_4 \partial_4,\
G=\partial _1,\ P=\partial_2,$
\\
& $\phantom{R_{\mathfrak h_{1.1}}\colon\ }
S=-x_2 \partial _1+\partial _3,\
T=x_1 \partial _2+x_3^2 \partial _3+(1-2 x_3 x_4) \partial _4$ \\
% subalgebra {T}
$\{G,\,P,\,S,\,D\}$ &$ R_{\mathfrak h_{1.2}}\colon\
D=-x_1 \partial _1+x_2 \partial _2-2 x_3 \partial _3+\partial _4,\
G=\partial _1,\
P=\partial _2,$
\\
& $\phantom{R_{\mathfrak h_{1.1}}\colon\ }
S=-x_2 \partial _1+\partial _3,\
T=x_1 \partial _2+x_3^2 \partial _3-x_3 \partial _4$
\\
% subalgebra {P}
$\{S,\,D,\,G,\,T\}$ &$ R_{\mathfrak h_{1.3}}\colon\
D=-2 x_1 \partial _1+\partial _2,\
G=e^{x_2} \partial _3,\
P=x_1e^{x_2}  \partial _3,$
\\
& $\phantom{R_{\mathfrak h_{1.1}}\colon\ } S=\partial_1,\
T=x_1^2 \partial_1-x_1 \partial_2+e^{-2 x_2} \partial_4$ \\
% subalgebra {S+T}
$\{G,\,P,\,S,\,D\}$ &$ R_{\mathfrak h_{1.4}}\colon\
D=-x_1 \partial_1+x_2 \partial_2-2 x_3 \partial_3+\partial_4,\
G=\partial _1,\
P=\partial _2,$\\
& $\phantom{R_{\mathfrak h_{1.1}}\colon\ } S=-x_2 \partial _1+\partial _3,\
T=x_1 \partial _2+\left(x_3^2-e^{-4 x_4}\right) \partial _3-x_3 \partial _4$ \\
% subalgebra {G+T}
$\{G,\,P,\,S,\,D\}$ &$R_{\mathfrak h_{1.5}}\colon\
D=-x_1 \partial_1+x_2 \partial_2-2 x_3 \partial_3+\partial_4,\
G=\partial_1,\
P=\partial_2,$
\\
& $\phantom{R_{\mathfrak h_{1.1}}\colon\ }
S=-x_2 \partial_1+\partial_3,\
T=-e^{-3 x_4} \partial_1+x_1 \partial_2+x_3^2 \partial_3-x_3 \partial_4$
\\
$\{G,\,P,\,S\}$ &$ R_{\mathfrak h_{2.1}}\colon\
D=-x_1\partial_1+x_2\partial_2-2 x_3 \partial _3,\
G=\partial_1,\
P=\partial_2,$
\\
&$\phantom{R_{\mathfrak h_{1.1}}\colon\ }
S=-x_2 \partial_1+\partial_3,\
T=x_3^2 \partial _3+x_1 \partial _2$ \\
% subalgebra {D,P}
$\{G,\,S,\,T\}$ &$ R_{\mathfrak h_{2.2}}\colon\
D=-x_1\partial_1-2 x_2 \partial_2+2 x_3 \partial_3,\
G=\partial_1,\
P=x_2 \partial_1,$\\&
$\phantom{R_{\mathfrak h_{1.1}}\colon\ }
S=\partial_2,\
T=x_1 x_2 \partial _1+x_2^2 \partial _2+ (1-2 x_2 x_3 ) \partial _3$ \\
% subalgebra {T,P}
$\{S,\,D,\,G\}$ &$ R_{\mathfrak h_{2.3}}\colon\
D=-2 x_1 \partial_1+\partial _2,\
G=e^{x_2} \partial_3,\
P= x_1 e^{x_2}\partial_3,\ S=\partial _1,\
T=x_1^2 \partial_1-x_1 \partial_2$ \\
% subalgebra {G,P}
$\{T,\,D,\,S\}$ &$ R_{\mathfrak h_{2.4}}\colon\
D=2 x_1 \partial_1+\partial_2,\
G=0,\
P=0,\
S=x_1^2 \partial_1+x_1 \partial _2+e^{2 x_2} \partial _3,\
T=\partial _1$ \\
% subalgebra {G+T,P}
$\{S,\,D,\,G\}$ &$ R_{\mathfrak h_{2.5}}\colon\
D=-2 x_1 \partial_1+\partial_2,\
G=e^{x_2} \partial _3,\
P= x_1 e^{x_2}\partial_3,$\\&$\phantom{R_{\mathfrak h_{1.1}}\colon\ }
S=\partial_1,\
T=x_1^2 \partial _1-x_1 \partial_2-e^{-2 x_2} \partial_3$ \\
% subalgebra {-G+T,P}
$\{S,\,D,\,G\}$ &$ R_{\mathfrak h_{2.6}}\colon\
D=-2 x_1 \partial _1+\partial _2,\
G=e^{x_2} \partial_3,\
P= x_1 e^{x_2}\partial_3,$\\&$\phantom{R_{\mathfrak h_{1.1}}\colon\ }
S=\partial_1,\
T=x_1^2 \partial _1-x_1 \partial _2+e^{-2 x_2} \partial_3$ \\
% subalgebra {D,S,T}
$\{G,\,P\}$ &$ R_{\mathfrak h_{3.1}}\colon\
D=-x_1 \partial _1+x_2 \partial _2,\
G=\partial_1,\
P=\partial_2,\
S=-x_2\partial_1,\
T=x_1 \partial_2$ \\
% subalgebra {D,T,P}
$\{G,\,S\}$ &$ R_{\mathfrak h_{3.2}}\colon\
D={-}x_1\partial_1{-}2 x_2 \partial_2,\
G=\partial_1,\
P=x_2 \partial_1,\
S=\partial_2,\
T=x_1 x_2 \partial _1+x_2^2 \partial _2$ \\
% subalgebra {D,G,P}
$\{S,\,T\}$ &$ R_{\mathfrak h_{3.3}}\colon\
D=-2 x_1 \partial_1+2 x_2 \partial_2,\
G=0,\
P=0,\
S=\partial_1,\  T=x_1^2 \partial _1+ (1-2 x_1 x_2 ) \partial_2$ \\
% subalgebra {T,G,P}
$\{S,\,D\}$ &$ R_{\mathfrak h_{3.4}}\colon\
D=-2 x_1 \partial_1+\partial_2,\
G=0,\
P=0,\
S=\partial _1,\
T=x_1^2 \partial_1-x_1 \partial _2$ \\
% subalgebra {S+T,G,P}
$\{S,\,D\}$ &$ R_{\mathfrak h_{3.5}}\colon\
D={-}2 x_1 \partial_1+\partial_2,\
G=0,\
P=0,\
S=\partial_1,\
T=\left(x_1^2-e^{{-}4 x_2}\right) \partial _1-x_1 \partial_2$
\\
% subalgebra {D,T,G,P}
$\{S\}$ &$ R_{\mathfrak h_{4.1}}\colon\
D=-2 x_1 \partial_1,\
G=0,\
P=0,\
S=\partial_1,\
T=x_1^2 \partial_1$
\\
\hline
\end{tabular}
%%\end{ruledtabular}
\caption{Realizations of the reduced special Galilei algebra ${{\rm A{G}}_3(1)}$.}
\label{tab:algag31}
\end{table*}

\section{Deformations of Galilei algebras} \label{sec_deformations}
Roughly speaking, deformation of a Lie algebra $\mathfrak{g}$ is the infinite set of Lie algebras that contains~$\mathfrak{g}$ and can be characterized
by common commutation relations parametrized by some continuous parameter.
Deformations of Lie algebras appear in mathematical theories when
it is necessary to classify non-isomorphic structures or to implement an analytical structure on the variety of such objects.
In physics deformations are used for quantization or for integration of theories with different symmetries into one model,
etc. In this paper we use the notion of \textit{one-parametric deformation} proposed by M.~Gerstenhaber~\cite{Gerstenhaber1964}.

Let $\mathfrak{g}=(V,[\ ,\ ])$ be a Lie algebra over $V$, then its \textit{deformation} is the one-parametric family of Lie algebras $\mathfrak{g}_q=(V,[\ ,\ ]_q)$, $q\in \mathbb{R}$, where
$[x ,y]_q=[x,y]+q\varphi_1(x,y)+q^2\varphi_2(x,y)+\cdots$,
and
$\varphi_i\colon V\times V \rightarrow V$ are bilinear, antisymmetric and satisfy the Jacobi identity.

Construction of all possible deformations of a given Lie algebra involves the Hochschild cohomology theory,
in particular, for the existence of infinitesimal deformation it is necessary that $\mathcal{H}^2(\mathfrak{g},\mathfrak{g})\ne 0$.
However in this section we avoid this complexity and construct explicitly several types of low-dimensional deformations
of Galilei algebras and two types of deformed Galilei algebras of arbitrary dimension.

\subsection{Deformation of the reduced classical Galilei algebra $\boldsymbol{{\rm A\bar{G}}_1(1)}$}\label{ssec_def_A_G_11}
To deform the Galilei algebra ${{\rm A\bar{G}}_1(1)}$ we use the mutually inverses of the contraction and deformation, namely it was shown in~\cite{Nesterenko1} that the three-dimensional Lie algebra $A_{3.1}$ (which is isomorphic to ${{\rm A\bar{G}}_1(1)}$) belongs to the orbit closure of the Lie algebra $A^\alpha_{3.4}$ defined by the commutation relations
\begin{equation*}
[e_1,e_3]=e_1,\quad [e_2,e_3]=\alpha e_2,\quad |\alpha| \leq  1,\quad \alpha \not= 0,1.
\end{equation*}
In particular, the In\"on\"u--Wigner contraction from $A^\alpha_{3.4}$ to $A_{3.1}$ can be realized by the contraction matrix
$\left(\begin{smallmatrix}
q & 0& 0\\
\tfrac{1}{1-\alpha} & 1 & 0\\
0&0 &q\\
\end{smallmatrix}\right),$
where $q$ is the contraction parameter.
Direct application of this contraction matrix allows us to construct the deformed Lie algebra
${{\rm A\bar{G}}^q_1(1)}=\langle P,\, G,\, T\rangle$
with the new $q$-parametrized product $[\cdot,\cdot]_q$ defined as follows
\begin{equation*}
[P,T]_q=qP,\quad [G,T]_q=P+\alpha qG,\quad q\in \R.
\end{equation*}

In much the same way from the contraction connecting $\mathfrak{sl}(2,{\mathbb R})$ and $A_{3.1}$ we can  construct the deformation $A\bar G^{\tilde q}_1(1)$ of the Galilei algebra ${{\rm A\bar G}_1(1)}$
with the $\tilde q$-parametrized product $[\cdot,\cdot]_{\tilde q}$ defined as follows
\begin{equation*}
[P,G]_{\tilde q}=-2{\tilde q}G,\quad
[P,T]_{\tilde q}=2{\tilde q}T,\quad
[G,T]_{\tilde q}=P,\quad
{\tilde q}\in \R.
\end{equation*}

Using the contraction matrix connecting the Lie algebras $A_{2.1}\oplus A_1$ and $A_{3.1}$ we  can  construct the deformation $A\bar G^{\hat q}_1(1)$ of the Galilei algebra ${{\rm A\bar G}_1(1)}$
with the ${\hat q}$-parametrized product $[\cdot,\cdot]_{\hat q}$ defined as follows
\begin{equation*}
[P,G]_{\hat q}={\hat q}P,\quad
[G,T]_{\hat q}=P,\quad
{\hat q}\in \R.
\end{equation*}

Note, that for $q=0$ the deformed algebra ${{\rm A\bar{G}}^q_1(1)}$ coincides with the initial algebra ${{\rm A\bar{G}}_1(1)}$
and for $\alpha=-1$ and $q\ne 0$ ${{\rm A\bar{G}}^q_1(1)}$ is isomorphic to the smallest Poincar\'e algebra $p(1,1)=\langle P_0,\, P_1,\, J_{01}\rangle$.

The appearance of the Poincar\'e algebra is expected and very natural as far as it was proven by In\"on\"u and Wigner~\cite{Inonu&Wigner1953} that Poincar\'e algebra contracts to the Galilei algebra,
what means that if the velocity of light is assumed to go to infinity, the relativistic mechanics `transforms' into the classical mechanics.

The next analogous connection is expected for the six-dimensional Poincar\'e algebra $p(1,2)$ and one of the ${{\rm A{G}}_3(1)}$ or ${{\rm A\bar{G}}_4(1)}$ Galilei algebras.

\subsection{Deformation of the classical Galilei algebra $\boldsymbol{{\rm AG}_1(1)}$}\label{ssec_def_AG_11}
Using the same approach as in previous section we deform the Galilei algebra ${{\rm A{G}}_1(1)}$,
but in this case we make profit from the contraction of the algebra $A^\beta_{4.8}$
($[e_2,e_3]=e_1,\, [e_1,e_4]=(1+\beta) e_1,\, [e_2,e_4]=e_2,\, [e_3,e_4]=\beta e_3,\, |\beta| \leq  1$)
to the algebra $A_{4.1}$ (which is isomorphic to ${{\rm A{G}}_1(1)}$ with the basis change~(\ref{baseAG11})).

For $\beta \not= 1$ we can realize this contraction by the matrix
$\left(\begin{smallmatrix}
q & 0 & 0 & 0\\
0 & q & 0 & 0\\
0 & 0 & 0 & \tfrac{q}{\beta-1}\\
0& -1 & 1 & 0\\
\end{smallmatrix}\right)$,
where $q$ is the contraction parameter.
Applying this contraction matrix we obtain the deformed Lie algebra
${{\rm A{G}}^q_1(1)}=\langle P,\, G,\, T,\, M \rangle$
with the following $q$-parametrized product $[\cdot,\cdot]_q$:
\begin{alignat*}{4}
&[P,G]_q=-M,\quad && [M,T]_q=\tfrac{1+\beta}{1-\beta} qM,\quad &&& \\
&[P,T]_q=\tfrac{1}{1-\beta}qP,\quad && [G,T]_q=P+\tfrac{\beta}{1-\beta} qG,\quad && q\in \R.&
\end{alignat*}

The deformed algebra ${{\rm A{G}}^0_1(1)}$ coincides with the initial algebra ${{\rm A{G}}_1(1)}$
and for $\beta=-\tfrac 12$ and $q\ne 0$ the deformed Galilei algebra ${{\rm A{G}}^q_1(1)}$ is isomorphic to the Galilei algebra ${{\rm A{G}}_2(1)}$.

Existence of contraction from the Lie algebra $\mathfrak{sl}(2,{\mathbb R})\oplus A_1$ to the algebra $A_{4.1}$ allows as to construct another physically interesting deformation ${\rm AG}^{\tilde q}_1(1)$ of the Galilei algebra ${{\rm AG}_1(1)}$
 with the $\tilde q$-parametrized product $[\cdot,\cdot]_{\tilde q}$ defined as follows
\begin{equation*}
[P,T]_{\tilde q}={\tilde q}T,\quad
[G,P]_{\tilde q}=M+{\tilde q}G,\quad
[G,T]_{\tilde q}=P,\quad
{\tilde q}\in \R.
\end{equation*}

The contraction  matrix connecting the Lie algebras $A_{2.1}\oplus A_{2.1}$ and $A_{4.1}$ is
$\left(\begin{smallmatrix}
-{\hat q}^3 & -{\hat q}^2 & -{\hat q} & 0\\
0 & 0 & 0 & {\hat q}\\
0 & {\hat q}^3 & 0 & -{\hat q}\\
0 & 0 & {\hat q}^2 & 0
\end{smallmatrix}\right)$.
From this contraction we get
one more important deformation $AG^{\hat q}_1(1)$ of the classical Galilei algebra ${{\rm AG}_1(1)}$
 with the $\hat q$-parametrized product $[\cdot,\cdot]_{\hat q}$ defined as follows
\begin{equation*}
[G,M]_{\hat q}={\hat q}M,\quad
[P,T]_{\hat q}=-{\hat q}M+{\hat q}^2P,\quad
[G,P]_{\hat q}=M,\quad
[G,T]_{\hat q}=P,\quad
{\hat q}\in \R.
\end{equation*}

\subsection{Deformation of the reduced extended Galilei algebras $\boldsymbol{{\rm A\bar G}_2(1)}$ and $\boldsymbol{{\rm A\bar G}_2(n)}$}
\label{ssec_def_AG_21}
In this and next section we use another approach that also allows us to construct deformations explicitly,
namely, we use the fact that the considered Galilei algebra is isomorphic to a representative of the infinite class of Lie algebras
from the classifications~\cite{mubarakzyanov1963.1,mubarakzyanov1963.2}.
It already was indicated in Section~\ref{sec_AG_21} that ${{\rm A\bar G}_2(1)}$ is isomorphic to $A^\beta_{4.8}$ for $\beta=-\tfrac 12$,
so we can use the parametrized commutation relations of $A^\beta_{4.8}$ and the isomorphism transformation~(\ref{baseAG21})
to construct the deformed Galilei algebra ${A \bar G^q_2(1)}$ with the product defined as
\begin{alignat*}{4}
&[T,G]_q=-P,\quad && [D,G]_q=G-2qG,\quad &&&\\
&[D,P]_q=-P-2qP,\quad && [D,T]_q=-2T,\quad && q\in \R.&
\end{alignat*}

The obtained deformation is also valid for the reduced extended Galilei algebra of any dimension, what can be proved by the induction method
and the deformed algebra ${A \bar G^q_2(n)}$ obey the standard relations~(\ref{comm_rel_JJJ})--(\ref{comm_rel_GJG}), (\ref{comm_rel_DTT})
together with the deformed relations
\begin{gather*}
 [D,G_i]_q=G_i -2qG_i,\quad [D,P_i]_q=-P_i-2qP_i,
\end{gather*}
where $i=1,2,\ldots,n$ and $q\in \R$.

\subsection{Deformation of the extended Galilei algebras $\boldsymbol{{\rm AG}_2(1)}$ and $\boldsymbol{{\rm AG}_2(n)}$}
In this case we make use of the series of five-dimensional Lie algebras $A^h_{5.30}$ with the commutation relations
\begin{alignat*}{4}
& [e_2,e_4]=e_1, \quad && [e_3,e_4]=e_2, \quad && [e_1,e_5]=(2+h)e_1,& \\
& [e_2,e_5]=(1+h)e_2, \quad && [e_3,e_5]=he_3,\quad  && [e_4,e_5]=e_4,&
\end{alignat*}
where $h\in \R$. The isomorphism ${{\rm AG}_2(1)}\sim A^{-2}_{5.30}$ was established in Section~\ref{sec_AG21} by the transformation~(\ref{baseAG21e}),
therefore the deformed Galilei algebra ${{\rm AG}^q_2(1)}$ with the parametrized product $[\cdot,\cdot]_q$ is defined by
\begin{alignat*}{4}
& [G,P]_q=M, \quad && [T,G]_q=-P,\quad && [D,G]_q=G, &\\
& [D,M]_q=qM,\quad && [D,P]_q=-P+qP, \quad && [D,T]_q=-2T+qT, &
\end{alignat*}
where $q\in \R$. By the induction method this deformation can be extended to the algebra ${{\rm AG}^q_2(n)}$, which obey the standard relations~(\ref{comm_rel_JJJ})--(\ref{comm_rel_DTT})
together with the deformed relations
\begin{gather*}
 [D,T]_q=-2T+qT,\quad [D,P_i]_q=-P_i+qP_i,\quad  [D,M]_q=qM,
\end{gather*}
where $i=1,2,\ldots,n$ and $q\in \R$.

\subsection{Varieties of the deformed Galilei algebras}
Deformations and contractions determine the partial order on the variety of Lie algebras of a fixed dimension, therefore, exploring the complete classification of contractions of low-dimensional Lie algebras~\cite{Nesterenko1}, we can make important conclusions on the structure of the  varieties of deformations of the Galilei algebras ${{\rm A\bar G}_1(1)}$, ${{\rm AG}_1(1)}$ and ${{\rm A\bar G}_2(1)}$.

The lowest Galilei algebra ${{\rm A\bar G}_1(1)}$ is three-dimensional and isomorphic to Heisenberg algebra, from the complete list of possible contractions contractions~\cite{Nesterenko1} it follows that ${{\rm A\bar G}_1(1)}$ can be deformed to any three-dimensional algebra except the Abelian algebra and the algebra $A_{3.3}$ ($[e_1,e_3]=e_1$, $[e_2,e_3]=e_2$).
So, all the differential equations (or their systems) invariant with respect to a three-dimensional Lie algebra (except $A_{3.3}$ and Abelian) in extreme case turn into Galilei-invariant theories.

In the case of four-dimensional Galilei algebras ${{\rm AG}_1(1)}$ and ${{\rm A\bar G}_2(1)}$ the situation is different.
The reduced extended Galilei algebra ${{\rm A\bar G}_2(1)}$ is isomorphic to the algebra $A^{-1/2}_{4.8}$, which belongs to one of the top levels of the contractions~\cite{Nesterenko1}, therefore the
variety of deformations of ${{\rm A\bar{G}}_2(1)}$ is
exhausted by the algebras constructed in Section~\ref{ssec_def_AG_21}.

Another four-dimensional Galilei algebra ${{\rm AG}_1(1)}$ is isomorphic to $A_{4.1}$ and has much more deformations, namely, classical Galilei algebra ${{\rm AG}_1(1)}$ can be deformed to all four-dimensional Lie algebras except the six cases: Abelian, $A^{1}_{4.2}$, $A^{a11}_{4.5}$, $A_{3.1}\oplus A_1$, $A_{3.3}\oplus A_1$ and $A_{2.1}\oplus 2A_1$.
Therefore, all but six theories with four-dimensional symmetry can be contracted into  ${{\rm AG}_1(1)}$-invariant theories.

\subsection{Generic realizations of the deformed Galilei algebras}
Using the method described in Section~\ref{sec_theory} we construct the generic realizations for the obtained deformations.

The generic realization of ${{\rm A\bar G}^q_1(1)}$ with the complement $\{P,\,G,\,T\}$ is
\begin{gather*}
e_1= \p_1,\quad
e_2= \p_2,\quad
e_3= (x_2+qx_1)\p_1+q\alpha x_2\p_2+\p_3,
\end{gather*}
where $q\in \R$ and $|\alpha| \leq  1$, $\alpha \not= 0,1$.

The generic realization of ${{\rm AG}^q_1(1)}$ with the complement $\{G,\,M,\,P,\,T\}$ is
\begin{align*}
e_1=&\p_2,\quad e_2=x_1\p_2+\p_3,\\
e_3=&q\tfrac{\beta}{\beta-1}x_1\p_1
+\left(\tfrac12 x_1^2-q\tfrac{\beta+1}{\beta-1}x_2\right)\p_2 +\left(x_1-\tfrac{q}{\beta-1}x_2\right)\p_3+\p_4,\quad
e_4= \p_1,
\end{align*}
where $q\in \R$ and $|\beta|\ \leq  1$, $\beta \not= 1$.

The generic realization of ${A\bar G^q_2(1)}$ with the complement $\{P,\,T,\,G,\,D\}$ is of the form
\begin{align*}
&e_1=\p_1,\quad e_2=\p_2,\quad e_3=-x_2\p_1+\p_3,\\
&e_4=(x_1+2\,x_1q)\p_1+2\,x_2\p_2+(-x_3+2\,x_3q)\p_3+\p_4,
\end{align*}
where $q\in \R$.

The generic realization of ${{\rm AG}^q_2(1)}$ with the complement $\{M,\,P,\,T,\,G,\,D\}$ is
\begin{align*}
& e_1= \p_1,\quad e_2=\p_2,\quad e_3=\p_3,\quad e_4=-x_2\p_1-x_3\p_2+\p_4,\\
& e_5=-qx_1\p_1+(x_2-qx_2)\p_2+(2x_3-qx_3)\p_3- x_4\p_4+\p_5,
\end{align*}
where $q\in \R$.

Note, that each of these realizations represents not only one Lie algebra, but the family of non-isomorphic Lie algebras at the same time.
As soon as the set of dependent and independent variables are fixed all the obtained realizations can be used for the construction of different types of equations and systems invariant with respect to the underlying symmetry groups,
e.g., with respect to the Galilei and Poincar\'e groups.

\section{Galilei-invariant differential equations}
\label{sec:galinvdifeq}
A number of well-known differential equations and systems appear to be Galilei-invariant, in this section we consider several particular examples of partial and ordinary differential equations that are invariant with respect to low-dimensional Galilei algebras
or their deformations.

\subsection{Burgers equation}
\label{ssec:burgers}
One of the most famous equations possessing the five-dimensional Galilei algebra is the Burgers equation~\cite{Burgers48}
\begin{gather*}
u_t+uu_x=\mu u_{xx},\quad \mu={\rm const}.
\end{gather*}
Being one of the simplest nonlinear $(1 + 1)$-dimensional evolution equations that is exactly solvable,
the Burgers equation has been used to describe many processes in fluid mechanics and a variety of other fields which seem to be rather disparate.  Its remarkable feature is that it can be transformed to the standard heat equation by means of the Hopf--Cole transformation \cite{Hopf50,Cole51}.
The maximal Lie symmetry algebra of the Burgers equation, that is a five-dimensional nonsolvable Lie symmetry algebra of the type $\mathfrak{sl}(2,\mathbb{R})\lsemioplus 2A_1$, was found as early as in 1965 by Katkov~\cite{Katkov}. This algebra is spanned by the generators
\begin{gather*}
\partial_t,\quad 2t\partial_t+x\partial_x-u\partial_u,\quad
t^2\partial_t+tx\partial_x+(x-tu)\partial_u,\quad\partial_x,\quad t\partial_x+\partial_u.
\end{gather*}
This set of differential operators can be constructed as the realization of Galilei algebra ${\rm A\bar{G}}_3(1)$
respective to the subalgebra $\langle D,\, -S\rangle$ and the complementary space $\{ P,\, G,\, -T\}$.

Note, that the subalgebra $\langle D,\, S\rangle$ does not belong to the list of the subalgebras obtained in~\cite{Fushchych&Barannyk&Barannyk}
and it is non-Abelian, so it should be automorphic to one of the cases $\langle D,\, T\rangle$ or $\langle D,\, P\rangle$.
In fact $\langle D,\, S\rangle$ is equivalent to $\langle D,\, T\rangle$ in the following way.
Basis elements of the first subalgebra are represented as
\begin{gather}\label{subalg_connection}
D=\alpha \tilde{T}+\beta \tilde{D},\quad S=\gamma \tilde{T}+\delta \tilde{D},
\end{gather}
where $\tilde{T}=a_1P+a_2T+a_3G+a_4D+a_5S$, $\tilde{D}=b_1P+b_2T+b_3G+b_4D+b_5S$ and
$\alpha,\beta,\gamma,\delta\in \R$, $\alpha\delta-\beta\gamma\ne0$,
the sets $a_i$ and $b_i$, $i=1,2,\ldots,5$  are the second and the fourth rows of an automorphism matrix.
Solving~(\ref{subalg_connection}) and using the last form of the automorphism transformation from Section~\ref{sec_AG_31}
we obtain one of solutions
$\alpha=\delta=0$, $\beta=-1$, $\gamma=\tfrac 1 {\alpha_{54}}$ and the automorphism is
$\left(\begin{smallmatrix}
0&0&-\alpha_{23}\alpha_{54}&0&0\\
0&0&0&0&\alpha_{54}\\
\alpha_{23}&0&0&0&0\\
0&0&0&-1&0\\
0&\frac{1}{\alpha_{54}}&0&0&0\\
\end{smallmatrix}\right),\text{ where }\alpha_{23}\alpha_{54}\ne0.
$

\subsection{Korteweg--de Vries and Kawahara equations}
\label{ssec:KdV}
The classical Korteweg--de Vries (KdV) equation and its generalizations model various physical systems, including
gravity waves, plasma waves and waves in lattices~\cite{Jeffrey}.
In this section we consider several types of the KdV-like equations,
that are invariant with respect to the three- or four-dimensional deformed Galilei algebras.

The prominent KdV equation,
\begin{equation*}
u_t+uu_x+u_{xxx}=0,
\end{equation*}
possesses the four-dimensional Lie invariance algebra of the type $A^{-2/3}_{4.8}$ with basis operators
\begin{equation*}
e_1=\partial_x,\quad e_2=\partial_t,\quad e_3=t\partial_x+\partial_u,\quad e_4=t\partial_t+\tfrac13x\partial_x-\tfrac23u\partial_u.
\end{equation*}

The \nth{5} order (quintic) KdV-equation  with constant coefficient,
\begin{equation*}
u_t+uu_x+ u_{xxxxx}=0,
\end{equation*}
possesses the four-dimensional Lie invariance algebra of the type $A^{-4/5}_{4.8}$ with basis operators
\begin{equation*}
e_1=\partial_x,\quad e_2=t\partial_x+\partial_u,\quad e_3=\partial_t,\quad e_4=t\partial_t+\frac15x\partial_x-\frac45u\partial_u.
\end{equation*}
Both these algebras are the particular cases of the deformed reduced extended Galilei algebra ${{\rm A \bar G}^q_2(1)}$
and their differential operators are equivalent to the realization respective to the one-dimensional subalgebra $\langle e_4\rangle$,
or, to the projection of the generic realization of the deformed algebra ${{\rm A \bar G}^q_2(1)}$ on the coordinates $(x_1, x_2, x_3)$
for the values $q=-\frac16$ and $q=-\frac{3}{10}$ respectively.

The Kawahara equation with time-dependent coefficient
\begin{equation*}
%&u_t+uu_x+\varepsilon t^\rho u_{xxx}=0,\\
u_t+uu_x+ \lambda t^\rho u_{xxx}+\varepsilon t^{\frac{5\rho+2}3}u_{xxxxx}=0
\end{equation*}
with $\lambda(\rho^2+\varepsilon^2)\neq0$ possesses the maximal Lie invariance algebra with basis operators~\cite{KPV2014}
\begin{equation}\label{sym_op_kdv}
 e_1=\partial_x,\quad e_2=t\partial_x+\partial_u,\quad
e_3=3t\partial_t+(\rho+1)x\partial_x+(\rho-2)u\partial_u,
\end{equation}
in particular, for $\varepsilon=0$ this equation becomes a variable-coefficient KdV equation of the form $u_t+uu_x+\lambda t^\rho u_{xxx}=0$ with the same symmetry.
According to the Mubarakzyanov's classification~\cite{mubarakzyanov1963.1} operators~(\ref{sym_op_kdv})
form the Lie algebra $A^{\alpha}_{3.4}$ for $\alpha=\frac{\rho-2}{\rho+1}$ ($e_3$ should be scaled by the factor $\frac{1}{\rho+1}$),
therefore this set of operators is equivalent to the generic realization of the deformed Galilei algebra
${{\rm A\bar G}^q_1(1)}$ for $q=1$ and $\alpha=\frac{\rho-2}{\rho+1}$.

Generalized KdV equation
\begin{equation*}
u_t=u^nu_{x}+\varepsilon u_{xxx},\quad n\neq0,1,\quad\varepsilon\neq0
\end{equation*}
possesses the symmetry algebra
\begin{equation*}
\langle\partial_t,\, \partial_x,\, 3nt\partial_t+nx\partial_x-2u\partial_u\rangle,
\end{equation*}
which is isomorphic to $A^{1/3}_{3.4}$ and equivalent to the generic realization of the deformed Galilei algebra
${{\rm A\bar G}^q_1(1)}$ for $q=1$ and $\alpha=\frac13$.

The 5-th order (quintic) KdV-equation  with time-dependent coefficient
\begin{equation*}
u_t+uu_x+ t^\rho u_{xxxxx}=0,\quad \rho\neq0
\end{equation*}
also has the Lie symmetry algebra that is isomorphic to $A^{\frac {\rho-4}{\rho+1}}_{3.4}$
\begin{equation*}
\langle\partial_x,\, t\partial_x+\partial_u,\, 5t\partial_t+(\rho+1)x\partial_x+(\rho-4)u\partial_u\rangle
\end{equation*}
and these operators are equivalent to the generic realization of the deformed Galilei algebra
${{\rm A\bar G}^q_1(1)}$ for $q=1$ and $\alpha=\frac {\rho-4}{\rho+1}$.

The classical Kawahara equation \begin{equation*}
u_t+uu_x+ \lambda u_{xxx}+\varepsilon u_{xxxxx}=0,\quad \lambda\varepsilon\neq0
\end{equation*} describes, in particular,  long waves in a shallow liquid under ice cover in the presence of tension or
compression~\cite{Marchenko}.
Its maximal Lie symmetry algebra
\[ \langle\partial_x,\, t\partial_x+\partial_u,\, \partial_t\rangle\]
 is isomorphic to  the Galilei algebra
${\rm A\bar G}_1(1)$ and presented operators coincide with the realiza\-tion~$R_{\mathfrak h_{0}}$, where the change of the variable $x_3\mapsto-x_3$
is performed.

Some other Galilei-invariant equations of Burgers and KdV types were constructed in~\cite{Boyko2001}.

\subsection{Reaction-diffusion equations}\label{ssec:Reaction-diffusion}

Reaction-diffusion equations and their systems are often used as model equations in mathematical biology, chemistry and physics. Thus,
in biology~\cite{murray2002} one can consider cells, bacteria, chemicals, animals and so on as particles
each of which moves around in a random way. Then, a regular motion of their group can be considered as a diffusion
process and often it is not a simple diffusion since there may be an interaction between particles. (1+1)-dimensional reaction-diffusion equations are widely used  for the description of global behavior
in terms of particle density or concentration. There are a number Galilei-invariant models among reaction-diffusion equations.

One of such examples is given by the reaction-diffusion equation
\begin{equation*}
u_t=(u^nu_x)_x+\varepsilon u^m,
\end{equation*}
where $n$ and $m$ are arbitrary constants
with \mbox{$n\varepsilon\neq0$,} \mbox{$m\neq0,1.$}
It was found in~\cite{Dorodnicyn}
that for $(n,m)\neq(-4/3,-1/3)$ this equation possesses the three-dimensional maximal Lie symmetry algebra with the basis operators
\begin{equation*}
e_1=\partial_t,\quad
e_2=\partial_x,\quad
e_3=2(1-m)t\partial_t+(1-m+n)x\partial_x+2u\partial_u,
\end{equation*}
which is transformed to the Lie algebra $A_{3.4}^{\frac{1-m+n}{2(1-m)}}$ by the isomorphism $\tilde e_3=\frac{1}{2(1-m)}e_3$.
Therefore, the reaction-diffusion equation is invariant with respect to the deformed reduced classical Galilei algebra ${\rm A\bar G}^q_1(1)$ for the deformation parameter $q=\frac{1-m+n}{2(1-m)}$ and the operators are equivalent to the generic realization of ${\rm A\bar G}^{q=\frac{1-m+n}{2(1-m)}}_1(1)$.

The maximal Lie invariance algebra of the  diffusion equation
\begin{equation*}
u_t=(u^nu_x)_x
\end{equation*}
with $n\neq0,\ -4/3$
is four-dimensional Lie algebra $A_{2.1}\oplus A_{2.1}$  spanned by the basis operators
\begin{equation*}
e_1=\partial_t,\quad
e_2=t\partial_t-\frac 1n u\partial_u,\quad
e_3=\partial_x,\quad
e_4=x\partial_x+\frac 2nu\partial_u.
\end{equation*}

So far as Lie algebra $A_{2.1}\oplus A_{2.1}$ belongs to the deformation variety of ${\rm AG}_1(1)$, the diffusion equation is invariant with respect to the deformed  classical Galilei algebra ${\rm AG}^{\hat q}_1(1)$ constructed in Section~\ref{ssec_def_AG_11}.

\subsection{Invariant systems of ODEs}
\label{ssec:Systems}
In this section we consider well-known systems of two second-order differential equations that are invariant with respect to the deformations of the three- and four-dimensional Galilei algebras.

%Ermakov systems have attracted attention in the last decades due to their important physical applications and mathematical properties as well.
%A central mathematical property of Ermakov systems is the existence of a constant of motion, the Ermakov invariant. The Ermakov invariant allows to construct a nonlinear superposition law linking the solutions of the equations of motion composing the Ermakov system. Ermakov systems have recently been of interest in diverse scenarios, such as accelerator physics,dielectric planar waveguides, cosmological models, analysis of supersymmetric families of Newtonian free damping modes, study of open fermionic systems, analysis of the propagation of electromagnetic waves in one-dimensional inhomogeneous media, algebraic approach to integrability of nonlinear systems, coupled linear oscillators, the semiclassical limit of quantum mechanics, supersymmetric quantum mechanics, computation of geometrical angles and phases for nonlinear systems, search for Noether and Lie symmetries, the possible linearization of the system, extension of the Ermakov system concept, the search for additional constants of motion and some discretizations of Ermakov systems.

Ermakov system was generalized in 1979~\cite{Ray} by Ray and Reid and since that time it is widely studied due to its importance in physics and mathematics such as existence of the Ermakov invariant (constant of motion), application to open fermionic systems, accelerator physics, quantum mechanics, etc.
Here we consider the generalized Ermakov system
\begin{equation*}
\ddot{x}=\frac{1}{x^3} F\left(\frac{y}{x}\right),\quad
\ddot{y}=\frac{1}{y^3} G\left(\frac{y}{x}\right).
\end{equation*}
The basis of the maximal Lie invariance algebra of this system for arbitrary $F$ and $G$  is given by the operators
\begin{equation*}
e_1=\partial_t, \quad e_2=2t\partial_t+x\partial_x+y\partial_y,\quad  e_3=t^2\partial_t+tx\partial_x+ty\partial_y,\end{equation*}
so the generalized Ermakov system is invariant with respect to the Lie algebra $\mathfrak{sl}(2,{\mathbb R})$ and the invariance operators are equivalent to the generic realization of $\mathfrak{sl}(2,{\mathbb R})$.
The Lie algebra $\mathfrak{sl}(2,{\mathbb R})$ belongs to the deformation variety of the Galilei algebra ${\rm A\bar G}_1(1)$, so the generalized Ermakov system is invariant with respect to the  deformed Galilei algebra ${\rm A\bar G}^{\tilde q}_1(1)$ defined in Section~\ref{ssec_def_A_G_11}.

Physical systems or models that deal with a force that points exactly toward (or away from) a force center usually involve the central force problem and appear in classical mechanics (force of the sun on a planet), atomic physics (motion of electron), etc.
Here we consider the two dimensional case of the central force problem
\begin{gather*}
{\ddot{\boldsymbol r}}+\left(\frac{\mu}{(x^2+y^2)^2}-\epsilon\right){\boldsymbol r}={\boldsymbol 0},\quad
\mu, \epsilon>0, \quad {\boldsymbol r}=(x,y).
\end{gather*}
This system can be rewritten via the polar coordinates in the form
\begin{gather*}
\ddot{r}-r\dot{\varphi}^2-\frac{\mu}{r^3}+\epsilon r=0,\quad
r\ddot{\varphi}+2\dot{r}\dot{\varphi}=0.
\end{gather*}
The maximal Lie invariance algebra of this system has the basis
\begin{gather*}
e_1=\partial_t,\quad
e_2=\partial_\varphi,\quad e_3=\sin(2\sqrt{\epsilon}t)\partial_t+\sqrt{\epsilon}\cos(2\sqrt{\epsilon}t)r\partial_r,\\ e_4=\cos(2\sqrt{\epsilon}t)\partial_t-\sqrt{\epsilon}\sin(2\sqrt{\epsilon}t)r\partial_r.
\end{gather*}
This basis forms the four-dimensional Lie algebra with the commutation relations
\begin{equation*}
[e_1,e_3]=2\sqrt{\epsilon}e_4,\quad
[e_1,e_4]=-2\sqrt{\epsilon}e_3,\quad [e_3,e_4]=-2\sqrt{\epsilon}e_1,
\end{equation*}
the isomorphism
$\tilde e_i=\frac 1{2\sqrt{\epsilon}}$ transforms this algebra to the $\mathfrak{sl}(2,{\mathbb R})\oplus A_1$,
and differential operators are equivalent to the realization of $\mathfrak{sl}(2,{\mathbb R})\oplus A_1$ with respect to the subalgebra $\langle e_2\rangle$.

As far as the Lie algebra $\mathfrak{sl}(2,{\mathbb R})\oplus A_1$ belongs to the deformation variety of the Galilei algebra ${\rm AG}_1(1)$, so the two-dimensional central forth problem invariant with respect to the  deformed Galilei algebra ${\rm AG}^{\tilde q}_1(1)$ with the $\tilde q$-parametrized product $[\cdot,\cdot]_{\tilde q}$ defined in Section~\ref{ssec_def_AG_11}.

Classical Kepler problem is one of the fundamental problems in classical mechanics and electrostatics,
in terms of polar coordinates it can be rewritten via the polar coordinates in the form of the second order system
\begin{gather*}
\ddot r-r\dot\theta^2+\frac{\mu}{r^2}=0,\quad
r\ddot\theta+2\dot r\dot \theta=0.
\end{gather*}
This system  is invariant with respect to the Lie algebra spanned by the operators
\begin{gather*}
e_1=\partial_t,\quad
e_2=\partial_\theta,\quad e_3=t\partial_t+\frac 23 r\partial_r.
\end{gather*}
These operators form the three-dimensional Lie algebra $A_{2.1}\oplus A_1$ and correspond to the generic realization, therefore, the classical Kepler problem is invariant with respect to the deformed Galilei algebra
${\rm A\bar G}^{\hat q}_1(1)$ defined in Section~\ref{ssec_def_A_G_11}.

\section{Conclusion}

It is well known that invariance principles allow us to explore the dynamical laws possessed by the physical systems, them also help us to restrict the possible form and to find these laws as well as to reduce some of their arbitrariness, etc. One of the fundamental symmetries in physics is the Galilei invariance and this symmetry underlies the Galilei relativity principle.

In this work we have considered five smallest Galilei algebras, notably the Lie algebras of dimensions three, four and five, in each of these cases we applied
an algebraic approach~\cite{shirokov2013} and constructed the complete set of in equivalent realizations. In particular all unfaithful realizations are new and lists of faithful realizations of two five-dimensional Galilei algebras are also new.
In addition we discussed possible deformations of the low-dimensional Galilei algebras and presented a number of deformations explicitly, some of them were constructed for arbitrary dimension of the algebra.
Finally, the generic realizations for several deformed Galilei algebras are constructed and it is shown that many of physically interesting partial differential equations and their systems are invariant with respect to the considered low-dimensional Galilei algebras and their deformations.
This motivates further investigation of the realizations of higher-dimensional Galilei algebras and their deformations.

\subsection*{Acknowledgments}
SP acknowledges the support of SGS15/215/OHK4/3T/14, project of the grant agency of the Czech Technical University in Prague.
OV and MN are grateful for the hospitality extended to them at the Department of mathematics, FNSPE, Czech Technical University in Prague.
The authors also would like to thank Vyacheslav Boyko, Roman Popovych and Anatoly Nikitin for constructive suggestions and comments.

%{\small \bibliographystyle{plain}
%\bibliography{realizations}
%}

\end{document}